\documentclass[amsmath,amssymb,reprint]{WileyMSP-template}

\usepackage{CJK}
\usepackage{amsmath}
\usepackage{ragged2e}
\usepackage{url}
\usepackage{cite}
\usepackage{color}
\usepackage[colorlinks=true,linkcolor=blue,citecolor=blue,anchorcolor=blue,urlcolor=blue]{hyperref}
\usepackage{lineno}
\usepackage{setspace}
\usepackage{array}
\usepackage{booktabs}
\usepackage{caption}
\usepackage{pdfpages}

\begin{document}
\begin{spacing}{1.1}
\begin{sloppypar}

\pagestyle{fancy}
\rhead{\includegraphics[width=2.5cm]{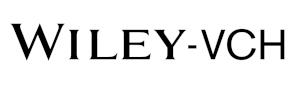}}

\title{Giant magnetocaloric effect in magnets down to the monolayer limit}

\maketitle


\author{Weiwei He}
\author{Yan Yin}
\author{Qihua Gong*}
\author{Richard F. L. Evans}
\author{Oliver Gutfleisch}
\author{Baixiang Xu}
\author{Min Yi*}
\author{Wanlin Guo*}



\begin{affiliations}
\small{W. He, Y. Yin, Q. Gong, M. Yi, W. Guo\\
State Key Lab of Mechanics and Control of Mechanical Structures \& Key Lab for Intelligent Nano Materials and Devices of Ministry of Education \& Institute for Frontier Science, Nanjing University of Aeronautics and Astronautics (NUAA), Nanjing 210016, China\\
Email Address: gongqihua@nuaa.edu.cn; yimin@nuaa.edu.cn; wlguo@nuaa.edu.cn

R. F. L. Evans\\
Department of Physics, The University of York, York YO105DD, United Kingdom\\

O. Gutfleisch, B. Xu\\
Institute of Materials Science, Technische Universit\"{a}t Darmstadt, Darmstadt 64287, Germany\\}

\end{affiliations}


\keywords{2D magnets, isothermal magnetic entropy change, adiabatic temperature change, specific cooling power, strain tunability}

\justifying  

\begin{abstract}
\noindent   
\large{Two-dimensional magnets could potentially revolutionize information technology, but their potential application to cooling technology and magnetocaloric effect (MCE) in a material down to the monolayer limit remain unexplored. Herein, we reveal through multiscale calculations the existence of giant MCE and its strain tunability in monolayer magnets such as CrX$_3$ (X\,=\,F,~Cl,~Br,~I), CrAX (A\,=\,O,~S,~Se;~X\,=\,F,~Cl,~Br,~I), and Fe$_3$GeTe$_2$. The maximum adiabatic temperature change ($\Delta T_\text{ad}^\text{max}$), maximum isothermal magnetic entropy change, and specific cooling power in monolayer CrF$_3$ are found as high as 11\,K, 35\,\textmu J\,m$^{-2}$\,K$^{-1}$, and 3.5\,nW\,cm$^{-2}$ under a magnetic field of 5\,T, respectively. A 2\% biaxial and 5\% $a$-axis uniaxial compressive strain can remarkably increase $\Delta T_\text{ad}^\text{max}$ of CrCl$_3$ and CrOF by 230\% and 37\% (up to 15.3 and 6.0\,K), respectively. It is found that large net magnetic moment per unit area favors improved MCE. These findings advocate the giant-MCE monolayer magnets, opening new opportunities for magnetic cooling at nanoscale.}

\end{abstract}

\section{Introduction}
Two-dimensional (2D) magnets, such as Cr$_2$Ge$_2$Te$_6$, CrI$_3$, VSe$_2$, Fe$_3$GeTe$_2$, VI$_3$~\cite{Gong2017Discovery,Huang2017Layer,Bonilla2018Strong,Deng2018Gate,
Tan2018Hard,Kong2019VI3}, have motivated numerous explorations of novel magnetic properties and their applications down to the monolayer limit over the past few years~\cite{Lado2017On,Chen2019Boosting,Pizzochero2020Magnetic,Lin2018Critical,
Huang2018Electrical,Niu2019Coexistence}.
With the additional spin degree of freedom and emergent phenomena in the 2D limit, 2D magnets have shown intriguing prospects in fields such as spin valves~\cite{Wang2020Electrically}, magnetic tunnel junctions~\cite{Watanabe2018Shape}, magnetic random access memory~\cite{Hou20192D,Zhang2020Memory}, and quantum computing~\cite{Liu2018Screening,Littlejohn2019Large}. These applications of 2D magnets favor the development of miniaturized spintronic and magnonic devices~\cite{Chumak2015Magnon,MacNeill2016Control,Sadovnikov2017Spin,Zhang2019Van} and could potentially revolutionize the next-generation of information storage/transport technologies~\cite{Burch2018Magnetism,Song2018Giant,Frisenda2018Recent,
Gibertini2019Magnetic}. 
Comparatively, the application of 2D magnets in cooling technology has been relatively underdeveloped compared to the tremendous efforts put into exploiting 2D magnets for information technology. Considering the flexibility of 2D magnets and the potential for mechanical deformation, their application in cooling technology is promising.

Cooling technology by using magnetic materials is intrinsically attributed to the magnetocaloric effect (MCE). 
MCE is a magneto-thermodynamic phenomenon. By exposing magnets to an external magnetic field, a reversible temperature and entropy change could be achieved to result in a targeted cooling or heating. MCE in bulk magnetocaloric materials have been demonstrated to enable highly efficient and environmentally friendly solid-state cooling technology that is a promising alternative to conventional gas-compression refrigerators~\cite{Omer2008Energy,Shen2009Recent,Gutfleisch2011Magnetic,
Gutfleisch2016Mastering,Li2020Understanding}. Despite this, studies on MCE in layered van der Waals (vdW) materials are still rare. Layered vdW magnets provide a platform for studying the thickness dependent MCE, which have enormous applications in low-dimensional magnetic refrigeration.

Recently, MCE in the bulk counterparts of 2D vdW magnets have been investigated, and the figures of merit for the MCE evaluation (e.g., maximum adiabatic temperature change $\Delta T_\text{ad}^\text{max}$, maximum magnetic entropy change $-\Delta S_\text{M}^\text{max}$) are reported~\cite{Liu2018Anisotropic,Yu2019Large,Liu2020Anisotropic,
Mondal2020Magnetic,liu2020Critical,Tran2022Insight}. 
For instance, by using the heat capacity data at an out-of-plane magnetic field up to 9\,T, $\Delta T_\text{ad}^\text{max}$ and $-\Delta S_\text{M}^\text{max}$ of bulk CrI$_3$ single crystals are estimated as 2.34\,K and 5.65\,J\,kg$^{-1}$\,K$^{-1}$, respectively. 
In addition, MCE in bulk CrI$_3$ single crystals is found to be anisotropic~\cite{Liu2018Anisotropic,Tran2022Insight} and the mechanism of this phenomenon is clarified theoretically, which depends on the anisotropic magnetic susceptibility and magnetization anisotropy~\cite{Tran2022Insight}. 
Similarly, for bulk CrBr$_3$ single crystals under an in-plane magnetic field of 5\,T, $\Delta T_\text{ad}^\text{max}$ and $-\Delta S_\text{M}^\text{max}$ are measured as around 2.37\,K and 7.2\,J\,kg$^{-1}$\,K$^{-1}$, respectively~\cite{Yu2019Large}. Later, bulk CrCl$_3$ are reported to have $-\Delta S_\text{M}^\text{max}$ of 14.6\,J\,kg$^{-1}$\,K$^{-1}$ under an in-plane magnetic field of 5\,T~\cite{Liu2020Anisotropic}. Furthermore, a magnetic field of 7\,T is shown to induce a large $\Delta T_\text{ad}^\text{max}$ of 6.2\,K and $-\Delta S_\text{M}^\text{max}$ of 19\,J\,kg$^{-1}$\,K$^{-1}$ in bulk CrCl$_3$~\cite{Mondal2020Magnetic}. In contrast to bulk CrI$_3$~\cite{Liu2018Anisotropic}, MCE in bulk CrCl$_3$ is isotropic~\cite{Mondal2020Magnetic}.
In contrast to typical magnetocaloric materials (such as La(Fe,Si)$_{13}$ family, Gd$_5$(Si,Ge)$_4$ family, and rare earth compounds)~\cite{Franco2018Magnetocaloric}, these vdW magnets exhibit superior magnetothermal properties and are suitable for use in low-temperature working environments up to 100\,K. Additionally, MCE in magnetic films is widely examined experimentally, but the film is of a thickness around 10$^{1}$--10$^{4}$\,nm~\cite{Teichert2015Structure,Dontgen2015Temp,Nguyen2021Magnetocaloric} and is much thicker than a monolayer layer. Thus, most of the current studies are restricted to bulk and film magnets and MCE in monolayer magnets remains to be explored.

In this work, we provide the new insight on the MCE of magnets down to the monolayer limit by a multiscale theoretical approach integrating \textit{ab-initio} calculations, atomistic spin simulations, and magnetocaloric thermodynamics. Specifically, monolayer magnets such as CrX$_3$ (X\,=\,F,~Cl,~Br,~I) and CrAX (A\,=\,O,~S,~Se;~X\,=\,F,~Cl,~Br,~I) are taken as model systems to explore their MCE. By using the magnetic parameters from \textit{ab-initio} calculations, atomistic spin model simulations are performed to determine the temperature dependent demagnetization curves, from which MCE is evaluated via the Maxwell relations. It is found that MCE indeed remains in these 2D magnets and can be remarkably tuned by strain. More importantly, giant MCE with a $\Delta T_\text{ad}^\text{max}$ around 15.3\,K is realized in magnets down to the monolayer limit.
These results provide theoretical guidance to probe MCE in 2D magnets, and could promote 2D magnets toward applications for cooling or thermal management in compact and miniaturized nanodevices.

\section{Results And Discussion}
\subsection{Strain-tunable magnetic properties}
\begin{figure}[!ht]
\centering
  \includegraphics[width=12cm]{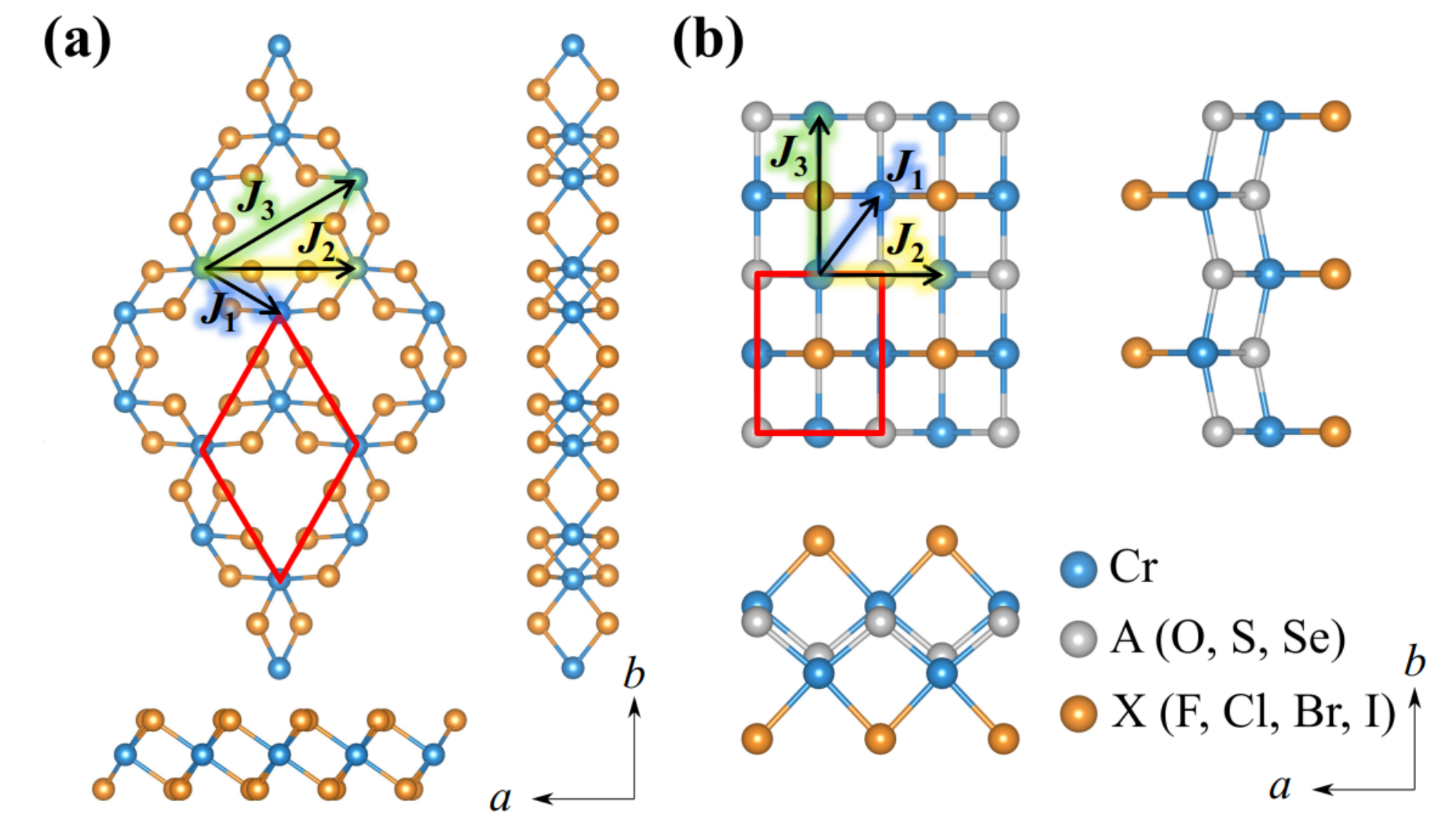}
  \caption{Crystal structures of (a) CrX$_{3}$ and (b) CrAX monolayers from the top and side views. The primitive cells of CrX$_3$ (two Cr and six X atoms) and CrAX (two Cr, two A atoms and two X atoms) have been indicated with red lines. The $c$ axis is perpendicular to the plane composed of $a$ axis and $b$ axis.  }
  \label{fig:structure}
\end{figure}

CrX$_3$ exhibits a rhombohedral lattice (space group $R \overline 3$), while the atomic arrangement of CrAX shows an orthorhombic structure (space group $Pmmn$). The top and side views of CrX$_3$ and CrAX structures are shown in Fig \ref{fig:structure}, in which the  associated primitive cells are highlighted with red lines. The optimized lattice parameters are summarized in Table \ref{tab:summary}, agreeing well with the previous works~\cite{Zhang2015Robust,Xu2021A,Guo2018Chromium,Han2020Prediction}. Exfoliation experiments of CrI$_3$~\cite{Huang2017Layer}, CrBr$_3$~\cite{Zhang2019Direct}, CrOCl~\cite{Zhang2019Magnetism}, CrOBr~\cite{Lee2021Magnetic}, and CrSBr~\cite{Lee2021Magnetic} have been carried out to show that the intrinsic ferromagnetism can maintain from the bulk to few-layer or even monolayer thickness.
In addition, for certain CrX$_3$ and CrAX which are not experimentally synthesized yet, their dynamic stability and existence possibility have been confirmed by a suite of theoretical studies~\cite{Zhang2015Robust,Liu2016Exfoliating,Rong2020Ferromagnetism,Xiao2020Modulating}. The calculated phonon dispersion spectra of CrAX are shown in Fig.~S1 (supporting information), proving the dynamic stability of these monolayers.

\begin{table*}[!b]
\centering
\caption{Magnetic properties and magnetocaloric parameters including optimized lattice constants (a and b, Å), net magnetic moment per primitive cell (M$_\text{tot}$, $\mu_\text{B}$), magnetic moment per Cr atom (M$_\text{Cr}$, $\mu_\text{B}$), net magnetic moment per unit area ($\bar{\text{M}}_\text{tot}$, \textmu A), easy or hard axis direction, magnetocrystalline anisotropy energy (MAE, meV/Cr), the nearest ($J_1$, meV), next-nearest ($J_2$, meV), and third-nearest ($J_3$, meV) neighboring exchange interactions, Curie temperature ($T_\text{c}$, K), maximum magnetic entropy change at 5\,T ($-\Delta S_\text{M}^\text{max}$, \textmu J\,m$^{-2}$\,K$^{-1}$) and maximum adiabatic temperature change at 5\,T ($\Delta T_\text{ad}^\text{max}$, K) for the 13 monolayers.}
\label{tab:summary}
\resizebox{\textwidth}{!}{
\renewcommand\arraystretch{1.5}

\begin{tabular} {m{1cm}<{\centering} m{0.9cm}<{\centering} m{0.9cm}<{\centering}  m{1cm}<{\centering} m{1cm}<{\centering} m{1cm}<{\centering} m{1.5cm}<{\centering} m{1cm}<{\centering}  m{1cm}<{\centering} m{1cm}<{\centering} m{1cm}<{\centering}  m{1.1cm}<{\centering}  m{1.8cm}<{\centering} m{1.8cm}<{\centering} }
\toprule
           &      &      & \multicolumn{5}{c}{Magnetic parameters} & \multicolumn{3}{c}{Exchange parameters} &     & \multicolumn{2}{c}{Magnetocaloric  parameters} \\
\cmidrule(lr){4-8}  \cmidrule(lr){9-11} \cmidrule(lr){13-14}        
Name & a & b & M$_\text{tot}$ & M$_\text{Cr}$ & $\bar{\text{M}}_\text{tot}$ & easy/hard axis & MAE    & $J_1$ & $J_2$ & $J_3$ & $T_\text{c}$ & $-\Delta S_\text{M}^\text{max}$ & $\Delta T_\text{ad}^\text{max}$  \\
\midrule
CrF$_3$ & \multicolumn{2}{c}{5.16} & 5.82 & 2.87 & 233.6 & $c$(E) & 0.122 & 3.81 & 0.15 & -0.03 & 21 & 35.04 & 10.98 \\
CrCl$_3$ & \multicolumn{2}{c}{6.05} & 5.68 & 2.94 & 166.5 & $c$(E) & 0.032 & 4.21 & 0.50 & -0.28 & 26 & 22.55 & 4.64 \\
CrBr$_3$ & \multicolumn{2}{c}{6.43} &5.72 & 3.01 & 147.9 & $c$(E) & 0.202 & 5.81 & 0.86 & -0.33 & 42 & 16.07 & 1.94 \\
CrI$_3$ & \multicolumn{2}{c}{7.00} & 5.74 & 3.10 & 125.3 & $c$(E) & 0.745 & 6.54 & 1.44 & -0.35 & 62 & 12.98 & 1.63 \\
CrOF & 3.09 & 3.88 & 6.02 & 3.18 & 466.2 & $c$(E) & 0.032 & 3.6 & 4.0 & 7.7 & 82 & 31.69 & 4.37 \\
CrOCl & 3.24 & 3.93 & 6.02 & 3.24 & 438.8 & $c$(E) & 0.022 & 3.5 & 2.3 & 7.9 & 67 & 31.63 & 4.75 \\
CrOBr & 3.36 & 3.94 & 6.05 & 3.30 & 423.9 & $b$(H) & 0.193 & 3.3 & 1.6 & 7.9 & 66 & 32.10 & 4.00  \\
CrSCl & 3.49 & 4.84 & 5.94 & 3.24 & 325.9 & $b$(H) & 0.007 & 13.6 & 11.7 & 3.6 & 161 & 12.65 & 2.11 \\
CrSBr & 3.59 & 4.83 & 5.95 & 3.27 & 318.2 & $b$(H) & 0.088 & 13.5 & 13.6 & 5.0 & 181 & 11.52 & 2.09 \\
CrSI & 3.76 & 4.81 & 5.98 & 3.32 & 306.6 & $b$(H) & 1.022 & 12.9 & 13.5 & 7.2 & 203 & 12.23 & 2.26 \\
CrSeCl & 3.58 & 5.17 & 6.02 & 3.35 & 301.5 & $a$(E) & 0.179 & 13.6 & 12.3 & -4.6 & 110 & 12.89 & 1.82 \\
CrSeBr & 3.68 & 5.12 & 6.02 & 3.37 & 296.1 & $a$(E) & 0.301 & 13.4 & 14.9 & -2.1 & 140 & 12.74 & 1.97 \\
CrSeI & 3.85 & 5.10 & 6.05 & 3.43 & 285.6 & $c$(E) & 0.921 & 13.6 & 16.4 & 1.4 & 192 & 11.67 & 2.15 \\
\bottomrule
\end{tabular}}
\end{table*}

The magnetic parameters of monolayer CrX$_3$ and CrAX are recorded in Table \ref{tab:summary}. Magnetism in these chromium-containing compounds originates from the incompletely filled electrons in $d$ orbitals of Cr ions. 
The net magnetic moment per primitive cell (M$_\text{tot}$) is about 6\,$\mu_\text{B}$, whereas the magnetic moment of Cr atoms (M$_\text{Cr}$) is clearly different in these monolayers. Magnetocrystalline anisotropy energy (MAE) the associated easy/hard axis directions are also summarized in Table \ref{tab:summary}. The monolayers with iodine atoms show strong magnetocrystalline anisotropy. For instance, CrI$_3$, CrSI, and CrSeI possess a large out-of-plane MAE of 0.745, 1.022, and 0.921\,meV, respectively, owing to the strong spin-orbit coupling (SOC) in iodine atoms. 
The out-of-plane energy of CrF$_3$, CrCl$_3$, CrBr$_3$, CrI$_3$, CrOF, CrOCl, and CrSeI is lower than the in-plane one, so an easy axis parallels to $c$ direction holds. Similarly, the easy axis of CrSeCl and CrSeBr is along the $a$ axis. Specially, since the energy in the $y$ direction is lager than that in other directions for CrOBr, CrSCl, CrSBr, and CrSI, the $b$ axis is regarded as the hard axis.

The exchange interaction parameters are also listed in Table \ref{tab:summary}. 
The positive values of exchange parameters $J_1$ and $J_2$ reflect the ferromagnetic coupling in monolayers. $J_1$ is increased by 2.73\,meV in CrX$_3$ with X varying from F to I, suggesting an enhancement in ferromagnetic coupling. Besides, $J_1$ and $J_2$ are one order of magnitude larger than $J_3$ in CrX$_3$, and are expected to dominate $T_{\text{c}}$ and the demagnetization behaviors. In contrast, for CrOX, $J_3$ is even larger than $J_1$ and $J_2$. 
This is possibly attributed to the different atomic structures for forming $J_2$ and $J_3$~\cite{Xiao2018Theoretical}. 
$J_3$ is formed by the super-exchange paths between two neighbor Cr atoms intermediated by an oxygen atom, i.e., Cr--O--Cr in 180$^\circ$. Whereas, $J_2$ is formed by two interactions Cr--O--Cr and Cr--X--Cr in 90$^\circ$~\cite{Zhang2019Super,Jiang2021Recent}.
Replacement of Cl with Br or I slightly changes $J_1$ and substantially enlarges $J_2$ and $J_3$ in CrSX and CrSeX.

\begin{figure}[!b]
\centering
  \includegraphics[width=12cm]{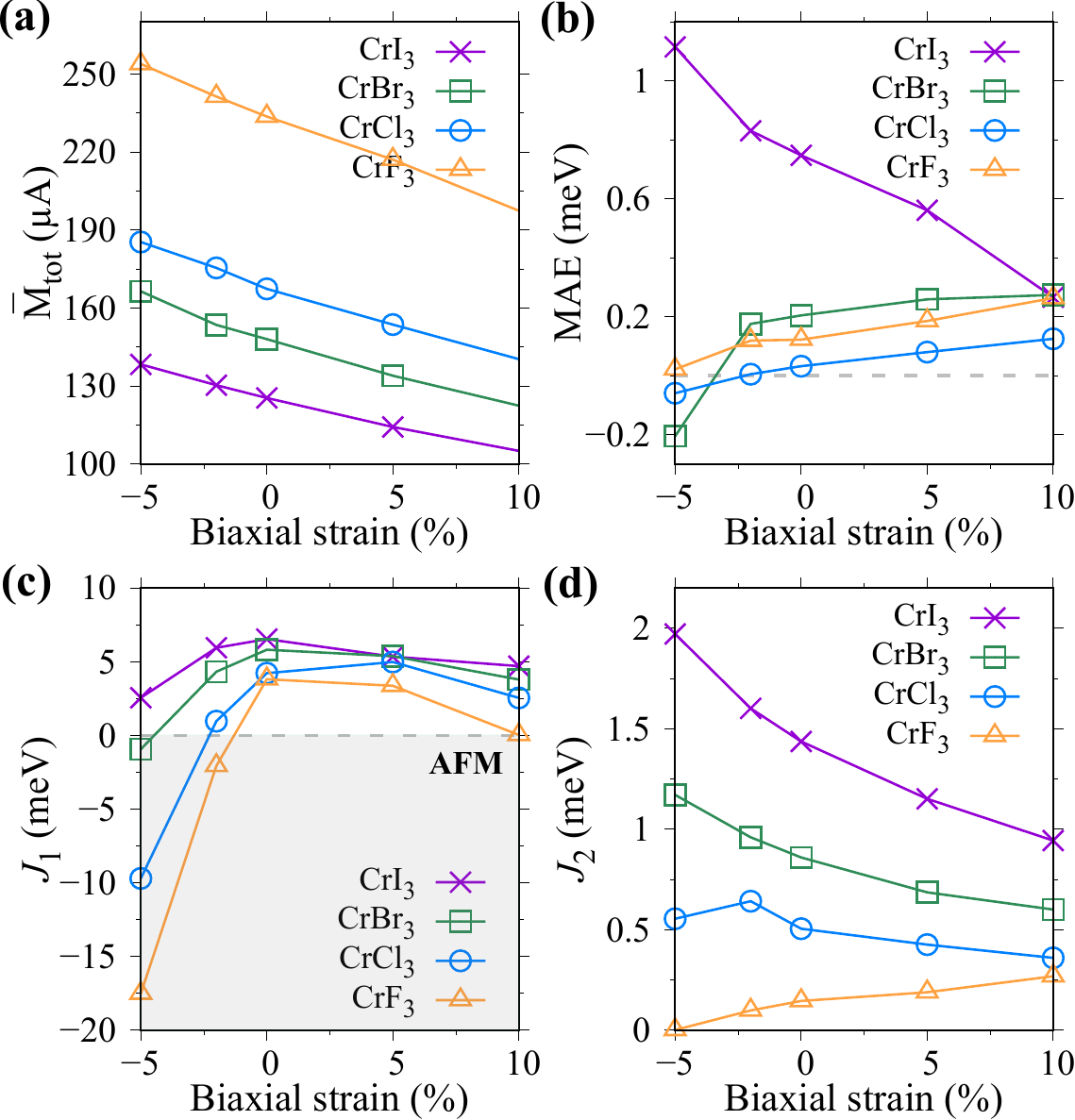}
  \caption{Strain-tunable magnetic properties of CrX$_3$: (a) net magnetic moment per unit area, (b) MAE, and exchange parameters (c) $J_1$ and (d) $J_2$.}
  \label{fig:mag}
\end{figure}

The influence of strain on magnetic properties are further examined. Biaxial strain is applied to CrX$_3$ and the results are depicted in Fig.~\ref{fig:mag}. 
The biaxial tensile strain slightly enhances M$_\text{Cr}$, but considerably increases the lattice area. Thus the slight increase of M$_\text{Cr}$ is canceled out and the net magnetic moment per unit area ($\bar{\text{M}}_\text{tot}$) is decreased, as seen in Fig.~\ref{fig:mag}a.
MAE of CrI$_3$ decreases with biaxial strain, while MAE of CrBr$_{3}$, CrCl$_{3}$, and CrF$_{3}$ increases with it, as shown in Fig.~\ref{fig:mag}b.
The impact of strain on exchange parameters $J_1$ and $J_2$ is demonstrated in Fig.~\ref{fig:mag}c and \ref{fig:mag}d, respectively. It is found that $J_1$ of CrX$_3$ is reduced by applying biaxial strain in spite of the tensile or compressive types. $J_1$ of CrBr$_3$, CrCl$_3$, and CrF$_3$ under --5\% compressive biaxial strain becomes negative, confirming that the modulation from ferromagnetic to antiferromagnetic coupling can be realized by applying compressive biaxial strain.
$J_2$ of CrF$_3$ and others (CrI$_3$, CrBr$_3$, CrCl$_3$) increases and decreases with the biaxial strain, respectively.
The effect of tensile and compressive uniaxial strains in CrAX is also investigated and the results are summarized in Figs.~S2 and S3. 
In addition, it is stated that~\cite{Sadhukhan2022Spin,Staros2022A} exchange parameters could be affected by finite temperature induced atom displacements. However, as shown in Fig.~S4, for 2D magnets whose MCE works at low temperatures, atom displacements are too small to notably affect magnetic properties, indicating the reasonable approximation of applying zero-K magnetic parameters in the classic spin Hamiltonians and the inconspicuous finite-temperature effect on MCE at low temperatures. Nevertheless, further in-depth studies are essential to calculate magnetic properties by considering atom displacements from vibration modes of 2D magnets if elevated temperatures are of interests.

\subsection{Demagnetization behavior at finite temperatures}

\begin{figure*}[!b]
\centering
  \includegraphics[width=18cm]{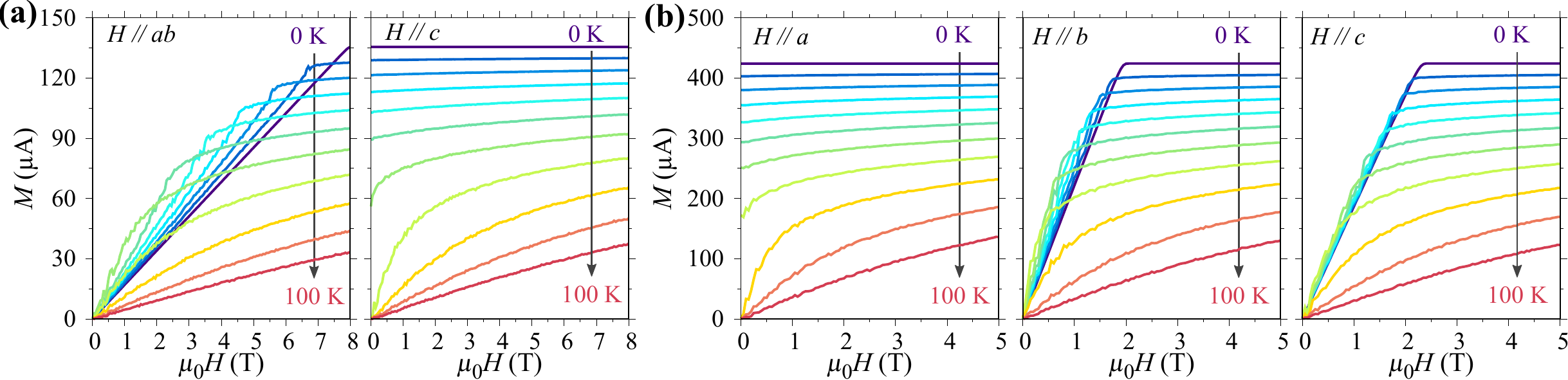}
  \centering
  \caption{Isothermal magnetization curves of (a) CrI$_3$ with field up to 8\,T and (b) CrOBr with field up to 5\,T applied in different directions. The magnetization ($M$) under different magnetic fields is defined as the net magnetic moment per unit area. The curves are displayed every 10\,K.}
  \label{fig:m-h}
\end{figure*}

The typical isothermal magnetization curves for CrI$_3$ and CrOBr at different temperatures are shown in Fig.~\ref{fig:m-h}. A striking distinction of saturation magnetization between CrI$_3$ and CrOBr can be noticed in Fig.~\ref{fig:m-h}a and \ref{fig:m-h}b.  For CrX$_3$ and CrAX with similar net magnetic moment, the monolayer with smaller lattice area possesses larger $\bar{\text{M}}_\text{tot}$ and thus higher saturation magnetization at the same temperature. The magnetization curves of other monolayers are shown in Figs.~S5--S8. 

Figure~\ref{fig:m-h} also indicates the anisotropy of the magnetization curve with respect to the direction of the applied magnetic fields, but the anisotropy is unconspicuous at elevated temperatures. Owing to the hexagonal structures of CrX$_3$, the demagnetization behavior is almost identical when the magnetic field is applied along $a$ or $b$ axis. The magnetization curves of CrI$_3$ are hard to saturate under an in-plane magnetic field, while they reach saturation easily under an out-of-plane one, as shown in Fig.~\ref{fig:m-h}a. 
This indicates $c$ as the easy axis at finite temperatures, agreeing with the high out-of-plane MAE of CrI$_3$ in Table
\ref{tab:summary}.
In contrast, as shown in Fig.~\ref{fig:m-h}b, CrOBr presents different magnetization curves in all the three crystallographic axes, owing to its in-plane tetragonal structure.
Our density functional theory (DFT) calculation with SOC predicts the highest energy with magnetization along $b$ axis ($E_b$) and thus $b$ as the hardest axis (i.e. $E_b>E_c>E_a$).
However, Fig.~\ref{fig:m-h}b indicates that magnetization along $c$ axis is even much harder to be demagnetized than that along $b$ axis.
This disagreement arises from the competition between the demagnetization energy in the atomistic spin model and the intrinsic MAE from DFT calculations. 
The minimization of demagnetization energy favors the in-plane alignment of magnetization, thus reducing $E_b$ by 0.201\,meV.
Since $E_b$ is only slightly lager than $E_c$ ($E_b-E_c\sim 0.193$\,meV), this energy reduction is enough to make $E_c$ exceed $E_b$ and thus $c$ as the harder axis.

Similar results are also summarized in  Figs.~S5-S8 for other monolayers.
The hard or easy axis determined by atomistic spin model simulations could differ from that by the DFT calculations. In addition to CrOBr, this phenomena exist as well in CrCl$_3$, CrOF, CrOCl, CrSCl, and CrSBr.
The distinction is mainly attributed to the demagnetization energy.
Our atomistic spin model includes the demagnetization field and thus the so-called extrinsic shape anisotropy, whereas DFT calculations only give the intrinsic MAE originated from SOC.

\subsection{Magnetocaloric effect and its strain tunability}

\begin{figure}[!b]
\centering
  \includegraphics[width=12cm]{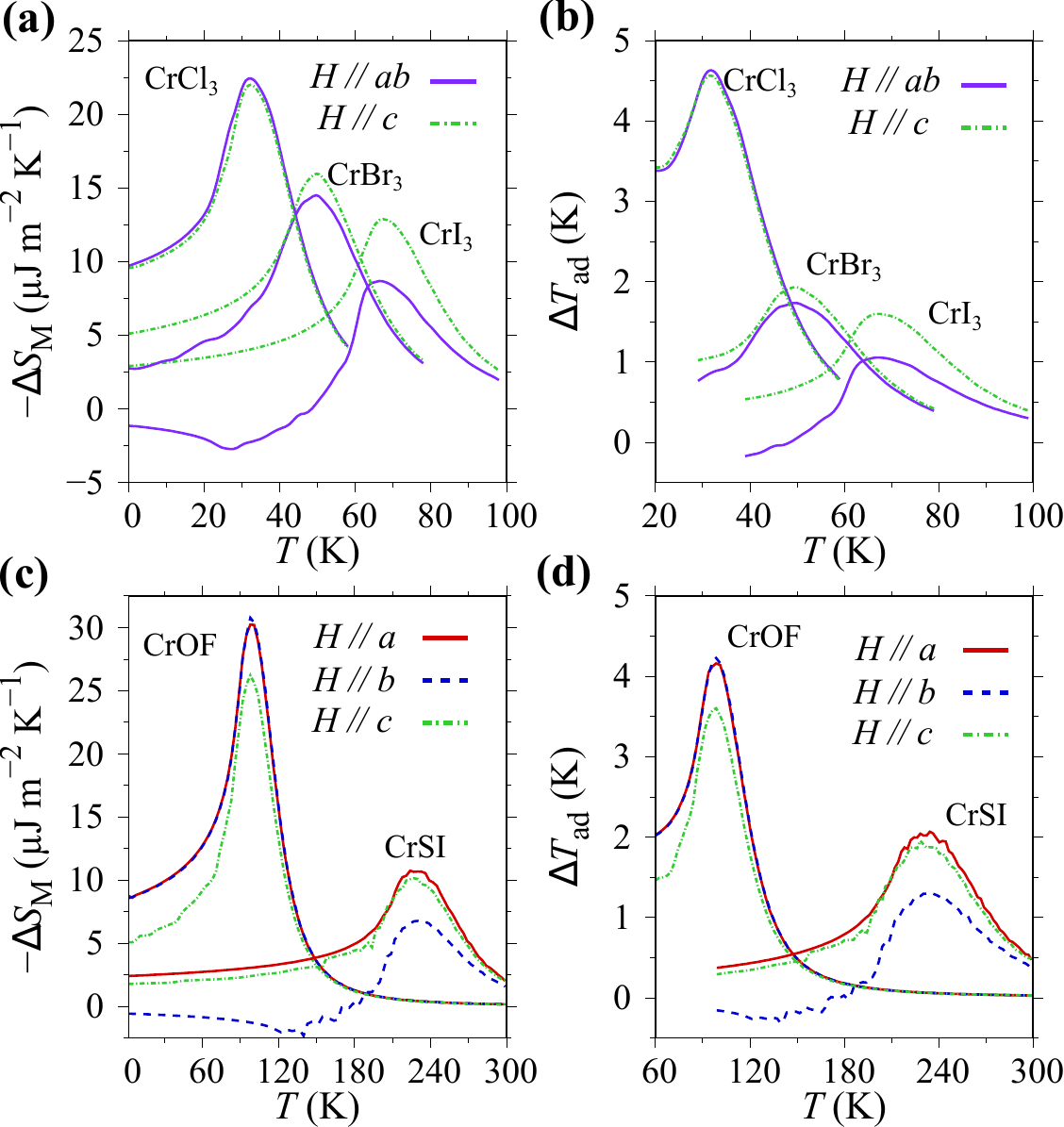}
  \caption{Temperature dependence of negative isothermal magnetic entropy change ($-\Delta S_\text{M}$) and the adiabatic temperature change ($\Delta T_\text{ad}$) under a magnetic field of 5\,T applied in different directions. (a), (b) Temperature dependence of $-\Delta S_\text{M}$ and $\Delta T_\text{ad}$ for CrCl$_3$, CrBr$_3$, and CrI$_3$. (c), (d) Temperature dependence of $-\Delta S_\text{M}$ and $\Delta T_\text{ad}$ for CrOF and CrSI.}
  \label{fig:st}
\end{figure}

First, we calculate the MCE of bulk CrI$_3$ and the results are shown in Fig.~S9. The calculated results match the experimental measurements~\cite{Liu2018Anisotropic} very well, confirming the reliability of the MCE calculation methodology that integrates \textit{ab-initio} calculations, Monte-Carlo simulations and magnetocaloric thermodynamics. Moreover, by evaluating MCE in monolayer Fe$_3$GeTe$_2$ from the experimental data~\cite{Deng2018Gate}, we find the consistence between experimental results and our theoretical predictions, as well as the indirect experimental evidence for the survive of MCE in monolayer Fe$_3$GeTe$_2$, as shown in Fig.~S10. The magnetocaloric parameters of monolayers including $\Delta T_\text{ad}^\text{max}$ and $-\Delta S_\text{M}^\text{max}$ under a magnetic field of 5\,T are summarized in Table \ref{tab:summary}.
It is found that CrF$_3$ outperforms other monolayers, with $\Delta T_\text{ad}^\text{max}$ and $-\Delta S_\text{M}^\text{max}$ exceeding 10\,K and 35\,\textmu J\,m$^{-2}$\,K$^{-1}$, respectively. The large $-\Delta S_\text{M}^\text{max}$ of CrF$_3$ is ascribed to the huge change of magnetization around $T_\text{C}$, as seen in  Fig.~S5a. The outstanding $\Delta T_\text{ad}^\text{max}$ of CrF$_3$ is determined by the smallest density and lowest specific heat capacity at low temperatures.
In addition, CrOF and CrOCl also exhibit excellent MCE with $\Delta T_\text{ad}^\text{max}$ and $-\Delta S_\text{M}^\text{max}$ as high as 4\,K and 31\,\textmu J\,m$^{-2}$\,K$^{-1}$, respectively. Since CrOF and CrOCl possess significantly large $\bar{\text{M}}_\text{tot}$ and relatively low $T_\text{C}$, their magnetization is more sensitive to the temperature change, thus boosting the $-\Delta S_\text{M}$ according to Equation.~\ref{eq:S}.
Besides, CrSX and CrSeX have a high $\Delta T_\text{ad}^\text{max}$ and $-\Delta S_\text{M}^\text{max}$ around 4\,K and 12\,\textmu J\,m$^{-2}$\,K$^{-1}$, respectively. 
Their $T_\text{C}$ between 100--200\,K makes them potentially applicable in magnetic refrigerants at medium temperatures.

\begin{figure}[!b]
\centering
  \includegraphics[width=12cm]{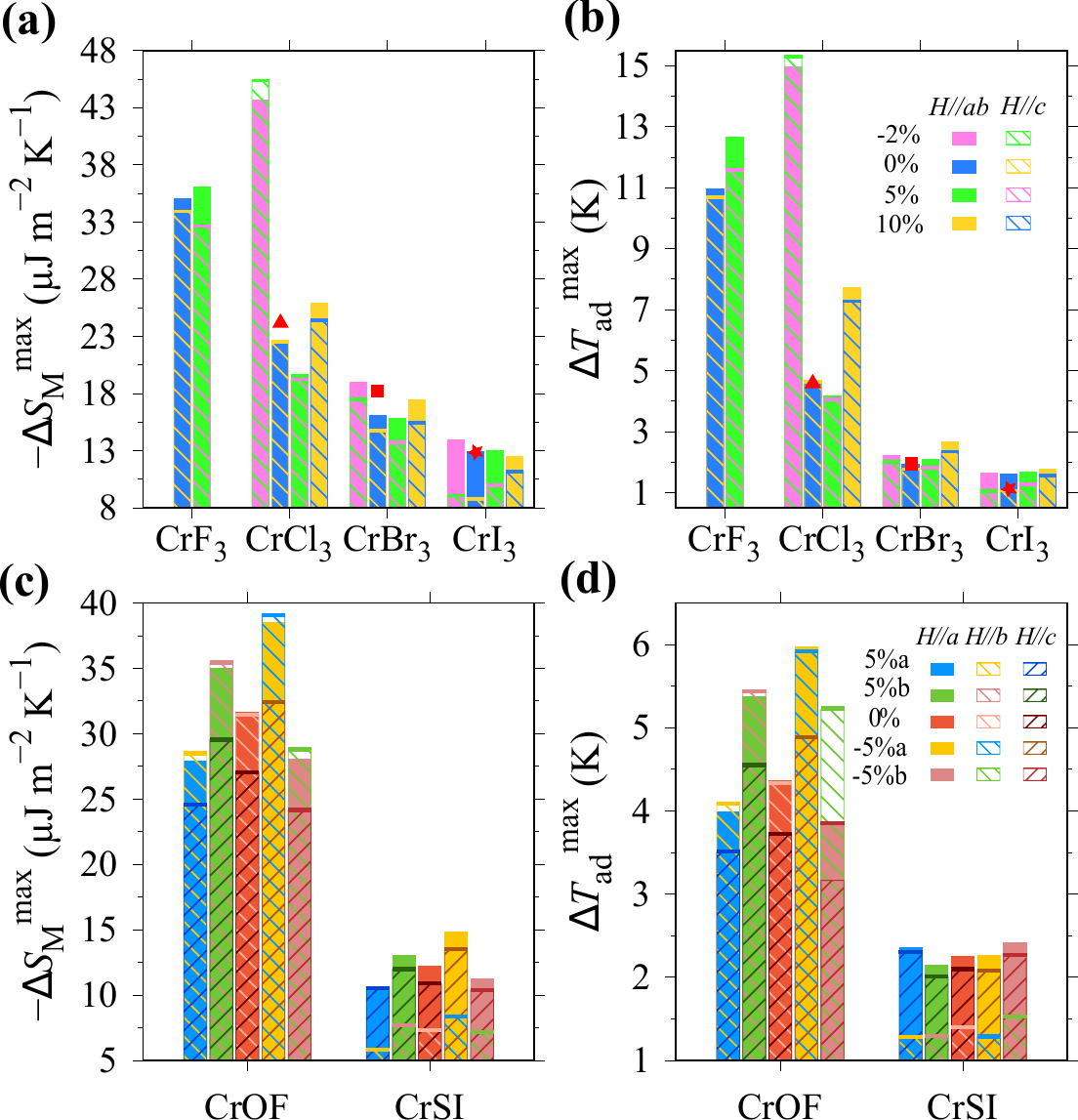}
  \caption{Strain-tunable MCE under a magnetic field of 5\,T applied in different directions. (a), (b) $-\Delta S_\text{M}^\text{max}$ and $\Delta T_\text{ad}^\text{max}$ at --2\%, 0\%, 5\%, and 10\% biaxial strain for CrX$_3$. (c), (d) $-\Delta S_\text{M}^\text{max}$ and $\Delta T_\text{ad}^\text{max}$ at 5\% $a$/$b$-axis tensile and compressive uniaxial strain for CrOF and CrSI. The experimental data of single crystal of bulk CrCl$_3$~\cite{Mondal2020Magnetic}, CrBr$_3$~\cite{Yu2019Large} and CrI$_3$~\cite{Liu2018Anisotropic} is symbolized by red triangles, squares, and pentagrams, respectively.}
  \label{fig:strain}
\end{figure} 

The temperature dependence of $-\Delta S_\text{M}$ and $\Delta T_\text{ad}$ under a magnetic field of 5\,T applied in different directions is shown in Fig.~\ref{fig:st}. 
$-\Delta S_\text{M}$ and $\Delta T_\text{ad}$ firstly increase with temperature and then reach their maximum around $T_\text{C}$. This maximum also increases with the applied magnetic field, as shown in Figs.~S11-S17.
Regardless of the directions of the applied magnetic fields at a fixed temperature, CrCl$_3$ exhibits an almost identical $-\Delta S_\text{M}$ and $\Delta T_\text{ad}$, as shown in Fig.~\ref{fig:st}a and \ref{fig:st}b.
This indicates that MCE of CrCl$_3$ is weakly direction dependent, agreeing with the experimental observations on the bulk counterparts~\cite{Liu2020Anisotropic}. The isotropic MCE behavior could be ascribed to the quite low MAE of CrCl$_3$ in Table \ref{tab:summary}.
In addition, CrOF shows isotropic in-plane MCE that is stronger than the out-of-plane one (Fig.~\ref{fig:st}c and \ref{fig:st}d). The stronger in-plane MCE is mainly owing to the much larger demagnetization energy induced by the highest $\bar{\text{M}}_\text{tot}$ of CrOF and thus much easier rotation of magnetization towards the $ab$ plane.
On the contrary, due to its large out-of-plane MAE, CrI$_3$ shows apparently anisotropic MCE. Specifically, $-\Delta S_\text{M}^\text{max}$ and $\Delta T_\text{ad}^\text{max}$ at 5\,T reach 12.98\,\textmu J\,m$^{-2}$\,K$^{-1}$ and 1.63\,K along the $c$ axis, which are 33\% and 35\% larger than that in the $ab$ plane, respectively.
Furthermore, there exists negative $ab$-plane MCE for CrI$_3$ at low temperatures. This is originated from the competition between the temperature dependence of magnetocrystalline anisotropy and magnetization~\cite{Liu2018Anisotropic}.
For CrI$_3$ at low temperatures, as the temperature increases, the out-of-plane MAE decreases much more rapidly than the magnetization, leading to a larger in-plane magnetization (Fig.~\ref{fig:m-h}). These explanations are also applicable to CrSI (Fig.~\ref{fig:st}c and \ref{fig:st}d). $-\Delta S_\text{M}^\text{max}$ and $\Delta T_\text{ad}^\text{max}$ of these 13 monolayers as a function of magnetic field ranging from 1 to 5\,T are summarized in Fig.~S18.

\begin{figure*}[!b]
\centering
  \includegraphics[width=18cm]{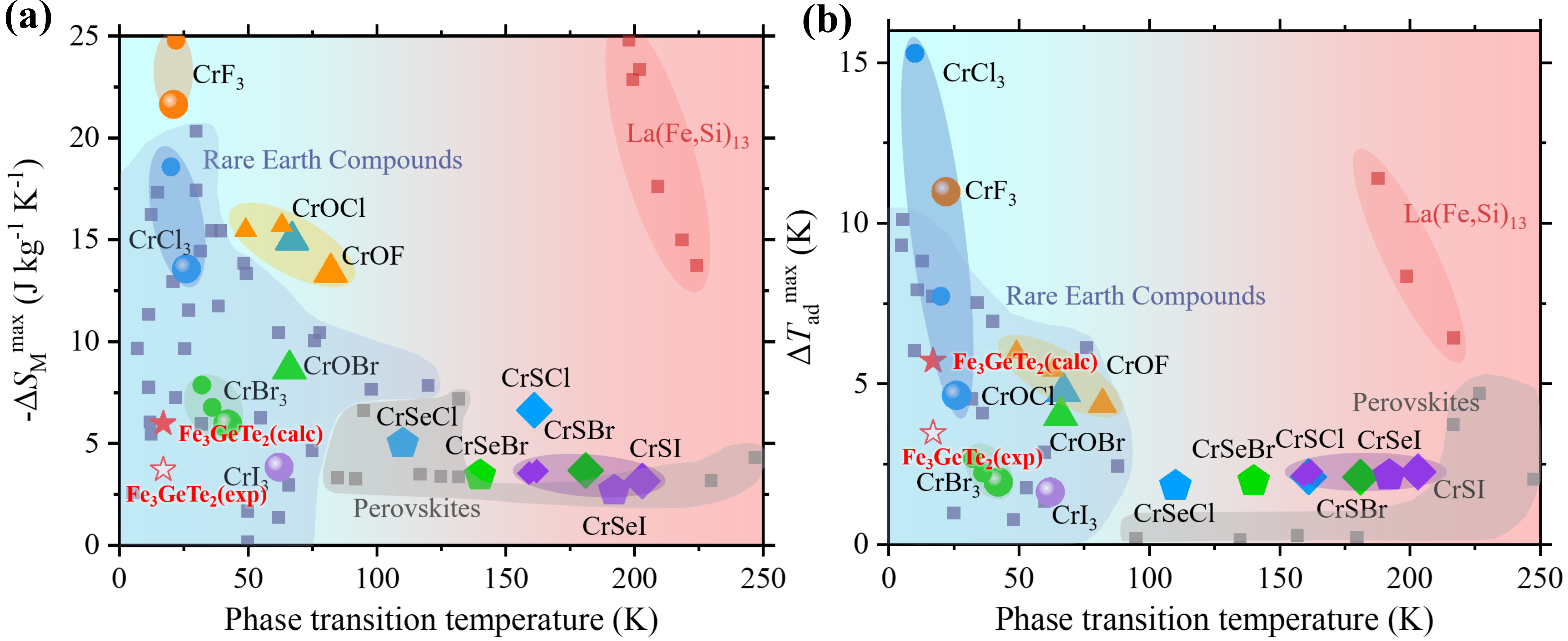}
  \caption{The theoretical and ideal comparison of (a) $-\Delta S_\text{M}^\text{max}$ and (b) $\Delta T_\text{ad}^\text{max}$ between the monolayers and classical bulk magnetocaloric materials (i.e., rare earth compounds~\cite{Midya2011Anisotropic,Franco2018Magnetocaloric,Zhang2019Review}, perovskites~\cite{Phan2004Large,Patra2009Multifunctionality,Aliev2018Magnetic,Gamzatov2018Correlation,Ram2018Review}, and La(Fe,Si)$_{13}$~\cite{Jia2009Magnetocaloric,Shen2009Recent,Franco2018Magnetocaloric} at 5\,T). CrX$_3$, CrOX, CrSX, CrSeX and Fe$_3$GeTe$_2$ are represented by circles, triangles, rhombuses, pentagons, and pentagrams, respectively. Monolayers containing F, Cl, Br, and I atoms are filled with orange, blue, green, and purple, respectively. The results with strain applied to CrF$_3$, CrCl$_3$, CrBr$_3$, CrOF, and CrSI are symbolized by smaller points. The experimental and calculated MCE of Fe$_3$GeTe$_2$ is from Fig.~S10~\cite{Deng2018Gate}.}
  \label{fig:compare}
\end{figure*}

To explore the effects of strain on MCE of monolayers, $-\Delta S_\text{M}^\text{max}$ and $\Delta T_\text{ad}^\text{max}$ with a magnetic field of 5\,T under different strains are shown in Fig.~\ref{fig:strain}. 
MCE is confirmed to weaken with decreasing layer thickness~\cite{Dontgen2015Temp}, while ignoring defects and vacancies of monolayers in our calculation allows for the evaluation of ideal magnetocaloric performance. It can be seen from Fig.~\ref{fig:strain}a and \ref{fig:strain}b that MCE in monolayers of CrX$_3$ (X\,=\,Cl, Br, I) matches well with that in the single-crystal bulk counterparts~\cite{Mondal2020Magnetic,Yu2019Large,Liu2018Anisotropic}, confirming that MCE can survive in the monolayer limit.
It is also found that a considerable enhancement in MCE of CrCl$_3$ is realized by a --2\% biaxial compressive strain, with $\Delta T_\text{ad}^\text{max}$ and $-\Delta S_\text{M}^\text{max}$ increased by 229\% and 101\%, respectively (Fig.~\ref{fig:strain}a and \ref{fig:strain}b).
This strain-induced enhancement is primarily a result of increasing $\bar{\text{M}}_\text{tot}$ and decreasing $T_\text{C}$ by the biaxial compressive strain.
For MCE in CrOF, a 5\% $a$-axis uniaxial compressive strain is revealed to induce the similar improvements in Fig.~\ref{fig:strain}c and \ref{fig:strain}d, with $\Delta T_\text{ad}^\text{max}$ and $-\Delta S_\text{M}^\text{max}$ of CrOF enhanced by 36.9\% and 23.3\%, respectively. 
Although the $b$-axis uniaxial tensile strain reduces $T_\text{C}$ as well, the decreased $\bar{\text{M}}_\text{tot}$ prevents the significant improvement of MCE.

\begin{figure}[!t]
\centering
  \includegraphics[width=12cm]{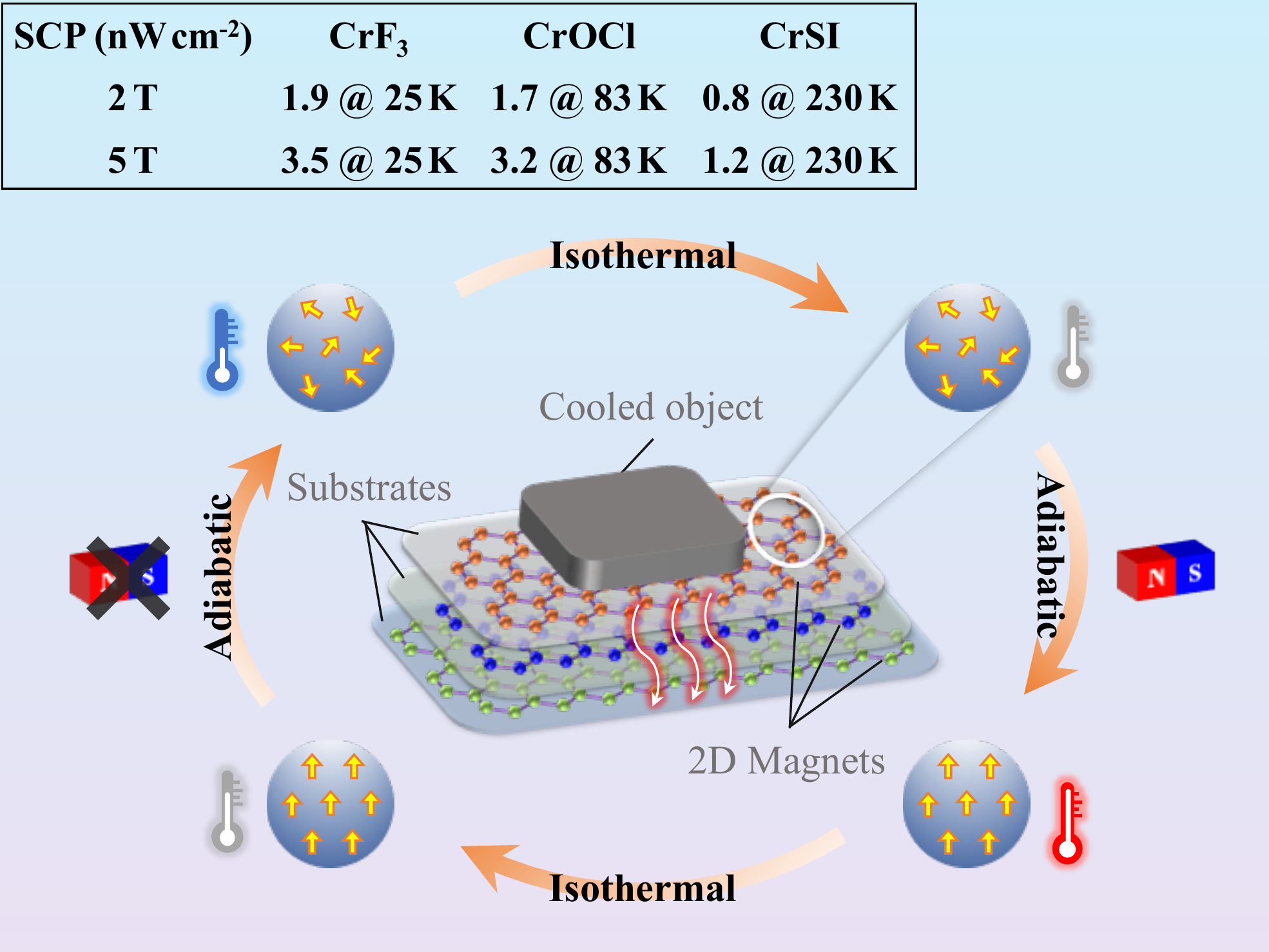}
  \caption{Schematic of refrigeration cycle using MCE of 2D magnets. SCP: specific cooling power.}
  \label{fig:cycle}
\end{figure}

Since 2D MCE is evaluated in terms of per area rather than per mass or volume, it should be converted to per kilogram before being compared to MCE of typical bulk materials. 1\,kg monolayer magnet corresponds to an area in the order of 10$^6$\,m$^2$ that is unattainable in reality. Here, we take an ideal assumption that 2D magnets could be infinitely large, so the comparison between 2D and bulk is only theoretically feasible. Based on this theoretical and ideal consideration, $-\Delta S_\text{M}^\text{max}$ and $\Delta T_\text{ad}^\text{max}$ of monolayers and bulk magnetocaloric materials are compared in Fig.~\ref{fig:compare}. 
The comparison between the monolayers and classical bulk MCE materials (e.g. conventional 1$^\text{st}$ and 2$^\text{nd}$ order materials, inverse 1$^\text{st}$ order materials) under a low magnetic field of 1 and 2\,T~\cite{Gottschall2019Making} is presented in  Fig.~S19.
It can be found from Fig.~\ref{fig:compare} that in terms of $-\Delta S_\text{M}^\text{max}$ and $\Delta T_\text{ad}^\text{max}$, MCE of monolayers is theoretically comparable to that of the bulk materials including typical rare-earth compounds~\cite{Midya2011Anisotropic,Franco2018Magnetocaloric,Zhang2019Review}, perovskites~\cite{Phan2004Large,Patra2009Multifunctionality,Aliev2018Magnetic,Gamzatov2018Correlation,Ram2018Review}, and La(Fe,Si)$_{13}$~\cite{Jia2009Magnetocaloric,Shen2009Recent,Franco2018Magnetocaloric}. 
We find that monolayer CrF$_3$ could work around 20--30\,K, which is close to the hydrogen liquefaction temperature. For the state of art of MCE in this temperature range, it is reported that bulk HoB$_2$ exhibits giant MCE with $-\Delta S_\text{M}^\text{max}$ \textasciitilde  40.1\,J\,kg$^{-1}$\,K$^{-1}$ and $\Delta T_\text{ad}^\text{max}$ \textasciitilde  12\,K for a field change of 5\,T~\cite{Castro2020Machine}. Therefore, it can be found from Fig.~\ref{fig:compare} that MCE of monolayer CrF$_3$ ($-\Delta S_\text{M}^\text{max}$ \textasciitilde  25\,J\,kg$^{-1}$\,K$^{-1}$, $\Delta T_\text{ad}^\text{max}$ \textasciitilde  11\,K under 5\,T) is comparable to that of bulk HoB$_2$ and rare-earth compounds near the liquid-hydrogen temperature.
Perovskites have the advantages of low price, good chemical stability, and widely tunable transition temperature~\cite{Ram2018Review,Zarkevich2020Viable}, while their MCE are unremarkable. 
The La-Fe-Si compounds with the first-order transition present the large entropy change near room temperature~\cite{Shen2009Recent,Ram2018Review}, but they suffer from lattice expansion and magnetic thermal hysteresis induced by the first-order transition.
In contrast, herein the 2D magnets not only possess giant MCE that is originated from the second-order phase transition without magnetic thermal hysteresis, but also are sufficiently flexible and deformable~\cite{Cantos2021Layer} to allow the further
tuning of MCE in 2D magnets by applying large strain. Therefore, mechanically robust 2D magnets could be promising candidates for the cooling of micro/nano devices.

As an illustration, Fig.~\ref{fig:cycle} depicts the possibly magnetic cooling enabled by 2D magnets. As a theoretical and ideal consideration, heat is assumed to be perfectly transferred outward from the cooled object when it contacts 2D magnets under an applied periodic magnetic field. For instance, using 2D CrF$_{3}$ as MCE material with an operation frequency of 1\,Hz, the ideal specific cooling power (SCP) around 25\,K is close to 1.9\,nW\,cm$^{-2}$ at 2\,T and 3.5\,nW\,cm$^{-2}$ at 5\,T. Under the same condition, the SCP of CrSI around 230\,K is 0.8\,nW\,cm$^{-2}$ at 2\,T and 1.2\,nW\,cm$^{-2}$ at 5\,T. The sequential stacking of 2D magnets with different working temperature windows could provide a cascade refrigeration cycle.

\section{Conclusion}

In summary, we confirm the survival of giant MCE and its strain tunability of magnets down to the monolayer limit incuding CrX$_3$ (X\,=\,F,~Cl,~Br,~I) and CrAX (A\,=\,O,~S,~Se;~X\,=\,F,~Cl,~Br,~I) by multiscale simulations.
It is found that CrF$_3$ exhibit excellent magnetocaloric performance at low temperatures owning to the smallest density and lowest specific heat capacity, with $\Delta T_\text{ad}^\text{max}$ and $-\Delta S_\text{M}^\text{max}$ up to 11\,K and 35\,\textmu J\,m$^{-2}$\,K$^{-1}$ under a magnetic field of 5\,T, respectively. Meanwhile, CrOF and CrOCl exhibit excellent MCE at medium temperatures with $\Delta T_\text{ad}^\text{max}$ and $-\Delta S_\text{M}^\text{max}$ as high as 4\,K and 31\,\textmu J\,m$^{-2}$\,K$^{-1}$, respectively. 
MCE of CrX$_3$ (X\,=\,F, Cl), CrAX (CrOF, CrOCl, CrSCl, CrSBr), and CrSeX  is isotropic, in-plane isotropic, and anisotropic, respectively. 
In addition, compressive strain could effectively enhance MCE by increasing $\bar{\text{M}}_\text{tot}$ and decreasing $T_\text{C}$. Particularly, a 2\% biaxial and 5\% $a$-axis uniaxial compressive strain can significantly increase $\Delta T_\text{ad}^\text{max}$ of CrCl$_3$ and CrOF to 15.3 and 6.0\,K, respectively.
Overall, these findings in our work extend the MCE reaserch to the monolayer limit.
Here, we reveal the consistency between experimental and calculated MCE for monolayer Fe$_3$GeTe$2$, as well as focus on the theoretical prediction of MCE and the associated mechanism in 2D magnets. The direct measurement of MCE in 2D magnets is currently challenging, but will be feasible as the measuring techniques are advanced.
Monolayer magnets with giant and strain-tunable MCE could enable their applications in the cooling of micro/nanoscale space and devices.

\section{Experimental Section}
\threesubsection{DFT Calculations}DFT calculations are performed using \textit{Vienna ab initio simulation page} (VASP)~\cite{Kresse1996Efficient,Kresse1999From} within the projector-augmented wave (PAW) method~\cite{Blochl1994Projector,Kresse1996Efficiency}. Exchange-correlation functional adopts Perdew-Burke-Ernzerhof (PBE) type~\cite{Perdew1996Generalized} in generalized gradient approximation (GGA). 
Considering that the calculated results of monolayer CrI$_3$ without $U$ are more consistent with the experimental values of bulk CrI$_3$~\cite{Dillon1965Magnetization}, $U$\,=\,0\,eV for CrX$_3$ is adopted~\cite{Zhang2015Robust}. The calculated magnetic properties of monolayer CrOCl and CrOBr with $U$\,=\,7\,eV for Cr are in better agreement with those in their bulk counterparts~\cite{Han2020Prediction}. The magnetic moment and Curie temperature of monolayer CrSBr calculated with $U$\,=\,3\,eV agree well with the experimental measurement ~\cite{LopezPaz2022Dynamic,Wilson2021Interlayer,Avsar2022Highly}. As a result, $U$\,=\,7\,eV for CrOX and $U=$\,3\,eV for CrSX and CrSeX are employed.
A 15-\AA -thick vacuum layer is set to prevent interaction between the periodic lattices along $c$ axis. The plane-wave cutoff energy is set to 500\,eV and the energy convergence criterion is 10$^{-6}$\,eV. The magnetic moment and MAE are evaluated in a primitive cell with a $k$-mesh of 15\,$\times$\,15\,$\times$\,1 and 27\,$\times$\,21\,$\times$\,1 generated by the Monkhorst-Pack scheme for CrX$_{3}$ and CrAX, respectively~\cite{Monkhorst1976Special}. MAE is calculated by the difference of total energies with the spin quantization axis aligned along different crystallographic axes. In detail, noncollinear non-self-consistent calculations with SOC are carried out by reading the converged charge densities from the spin-polarized self-consistent calculations. The phonon calculations are performed by the the finite displacement method in a 3\,$\times$\,3\,$\times$\,1 supercell~\cite{Togo2015First,Chaput2011Phonon,Togo2010First}.

In order to calculate the temperature dependent demagnetization curves of 2D magnets, we utilize the atomistic spin model that is based on the classic spin Hamiltonian~\cite{Skubic2008A}, i.e.,
\begin{equation}
\begin{aligned}
H &= E_0 - \frac{1}{2} J_1\sum_N \mathbf s_i \cdot \mathbf s_j - \frac{1}{2} J_2\sum_{NN} \mathbf s_i \cdot \mathbf s_j - \frac{1}{2} J_3\sum_{NNN} \mathbf s_i \cdot \mathbf s_j \\
&- k_i\sum_i(\mathbf s_i \cdot \mathbf e_i)^2- \sum_i \mu _s\mathbf s_i \cdot (\mathbf{H_{app}} + \mathbf{H_{dp}} ) ,
\end{aligned}
\label{eq:H}
\end{equation}
in which $E_0$ is the energy without spin contribution, $\mathbf s_i$ the unit vector representing the atomistic spin direction at atom $i$, $k_i$ the magnetocrytalline anisotropy energy of atom $i$, and $\mathbf e_i$ the easy axis vector. $\mu _s$ is the magnetic moment of atom $i$. $\mathbf{H_{app}}$ and $\mathbf{H_{dp}}$ are the external and dipole magnetic fields, respectively. $J_1$, $J_2$, and $J_3$ are the nearest-neighbour (NN), next-NN, and third-NN exchange interaction parameters, respectively.
These exchange parameters are derived by solving equations based on total energies of ferromagnetic and different antiferromagnetic configurations. 

\threesubsection{Atomistic Spin Simulations}After the parameters in Equation.~\ref{eq:H} have been obtained from \textit{ab-initio} calculations, the parameterized atomistic spin model is used to perform simulations by using VAMPIRE~\cite{Evans2014Atomistic}. The Monte Carlo method is adopted to acquire $T_{\text{c}}$. 
After executing 10,000 steps at each temperature, the system of 50$\times$\,50$\times$\,1 unit cells with in-plane periodic boundary conditions reaches equilibrium. Then a statistical average is taken over by further 10,000 steps to extract the mean magnetization. The temperature dependent magnetization curves are calculated by using the spin dynamics approach and the Heun integration scheme. The demagnetization field induced by the atomistic spins themselves is also included. The external magnetic field is gradually increased to 8\,T with a incremental step of 0.02\,T.

\threesubsection{Magnetocaloric Thermodynamics}The figures of merit for MCE are related to thermodynamics. The $\Delta S_\text{M}$ can be calculated from the Maxwell relation~\cite{Pecharsky1999Magnetocaloric,Gschneidner2000Magnetocaloric}, i.e.,
\begin{equation}
\Delta S_\text{M} = \int_0^H\left(\frac{\partial S}{\partial H}\right)_T\text{d}H = \mu_0\int_0^H\left(\frac{\partial M}{\partial T}\right)_H\text{d}H,
\label{eq:S}
\end{equation}
where $S$ is entropy, $M$ is magnetization, and $\mu_0$ is vacuum permeability. The degree of disorder in the magnetic moment distribution decreases as the magnetic field increases. So $\Delta S_\text{M}$ turns to a negative value. Thus, usually $-\Delta S_\text{M}$ is taken as one of the features to measure MCE. Similarly, $\Delta T_\text{ad}$ could demonstrate MCE directly, which can be represented as
\begin{equation}
\begin{aligned}
\Delta T_\text{ad} = -\mu_0\int_0^H \frac{T}{\rho c_\text{p}}\left(\frac{\partial S}{\partial H}\right)_{T}\text{d}H = -\mu_0\int_0^H \frac{T}{\rho c_\text{p}}\left(\frac{\partial M}{\partial T}\right)_{H}\text{d}H,
\end{aligned}
\label{eq:T}
\end{equation}
where $\rho$ is the density, and $c_\text{p}$ is the specific heat capacity which is acquired by phonon calculations.

\medskip
\noindent
\textbf{Supporting Information} \par 
\noindent
Supporting Information is available from the Wiley Online Library or from the author.

\medskip
\noindent
\textbf{Acknowledgements} \par
\noindent
The authors acknowledge the support from the National Natural Science Foundation of China (12272173, 11902150), the 15th Thousand Youth Talents Program of China, the Research Fund of State Key Laboratory of Mechanics and Control of Mechanical Structures (MCMS-I-0419G01 and MCMS-I-0421K01), the Fundamental Research Funds for the Central Universities, a project Funded by the Priority Academic Program Development of Jiangsu Higher Education Institutions, the European Research Council (ERC) under the European Unions Horizon 2020 research and innovation programme (No 743116), the German Research Foundation (DFG YI 165/1-1 and DFG XU 121/7-1).

\medskip
\noindent
\textbf{Conflict of Interest} \par 
\noindent
The authors declare no conflict of interest.


%
\bibliographystyle{MSP}
\bibliography{References}

\begin{thebibliography}{10}
\providecommand{\url}[1]{\texttt{#1}}
\providecommand{\urlprefix}{URL }

\bibitem{Gong2017Discovery}
C.~Gong, L.~Li, Z.~Li, H.~Ji, A.~Stern, Y.~Xia, T.~Cao, W.~Bao, C.~Wang,
  Y.~Wang, et~al.,
\newblock \emph{Nature} \textbf{2017}, \emph{546}, 7657 265.

\bibitem{Huang2017Layer}
B.~Huang, G.~Clark, E.~Navarro-Moratalla, D.~R. Klein, R.~Cheng, K.~L. Seyler,
  D.~Zhong, E.~Schmidgall, M.~A. McGuire, D.~H. Cobden, et~al.,
\newblock \emph{Nature} \textbf{2017}, \emph{546}, 7657 270.

\bibitem{Bonilla2018Strong}
M.~Bonilla, S.~Kolekar, Y.~Ma, H.~C. Diaz, V.~Kalappattil, R.~Das, T.~Eggers,
  H.~R. Gutierrez, M.~H. Phan, M.~Batzill,
\newblock \emph{Nature Nanotechnology} \textbf{2018}, \emph{13}, 4 289.

\bibitem{Deng2018Gate}
Y.~Deng, Y.~Yu, Y.~Song, J.~Zhang, N.~Z. Wang, Z.~Sun, Y.~Yi, Y.~Z. Wu, S.~Wu,
  J.~Zhu, J.~Wang, X.~H. Chen, Y.~Zhang,
\newblock \emph{Nature} \textbf{2018}, \emph{563}, 7729 94.

\bibitem{Tan2018Hard}
C.~Tan, J.~Lee, S.-G. Jung, T.~Park, S.~Albarakati, J.~Partridge, M.~R. Field,
  D.~G. McCulloch, L.~Wang, C.~Lee,
\newblock \emph{Nature Communications} \textbf{2018}, \emph{9}, 1 1554.

\bibitem{Kong2019VI3}
T.~Kong, K.~Stolze, E.~I. Timmons, J.~Tao, D.~Ni, S.~Guo, Z.~Yang, R.~Prozorov,
  R.~J. Cava,
\newblock \emph{Advanced Materials} \textbf{2019}, \emph{31}, 17 1808074.

\bibitem{Lado2017On}
J.~L. Lado, J.~Fern{\'{a}}ndez-Rossier,
\newblock \emph{2D Materials} \textbf{2017}, \emph{4}, 3 035002.

\bibitem{Chen2019Boosting}
S.~Chen, C.~Huang, H.~Sun, J.~Ding, P.~Jena, E.~Kan,
\newblock \emph{The Journal of Physical Chemistry C} \textbf{2019}, \emph{123},
  29 17987.

\bibitem{Pizzochero2020Magnetic}
M.~Pizzochero, R.~Yadav, O.~V. Yazyev,
\newblock \emph{2D Materials} \textbf{2020}, \emph{7}, 3 035005.

\bibitem{Lin2018Critical}
G.~Lin, X.~Luo, F.~Chen, J.~Yan, J.~Gao, Y.~Sun, W.~Tong, P.~Tong, W.~Lu,
  Z.~Sheng, et~al.,
\newblock \emph{Applied Physics Letters} \textbf{2018}, \emph{112}, 7 072405.

\bibitem{Huang2018Electrical}
B.~Huang, G.~Clark, D.~R. Klein, D.~MacNeill, E.~Navarro-Moratalla, K.~L.
  Seyler, N.~Wilson, M.~A. McGuire, D.~H. Cobden, D.~Xiao, et~al.,
\newblock \emph{Nature Nanotechnology} \textbf{2018}, \emph{13}, 7 544.

\bibitem{Niu2019Coexistence}
B.~Niu, T.~Su, B.~A. Francisco, S.~Ghosh, F.~Kargar, X.~Huang, M.~Lohmann,
  J.~Li, Y.~Xu, T.~Taniguchi, et~al.,
\newblock \emph{Nano Letters} \textbf{2019}, \emph{20}, 1 553.

\bibitem{Wang2020Electrically}
H.~Wang, J.~Qi, X.~Qian,
\newblock \emph{Applied Physics Letters} \textbf{2020}, \emph{117}, 8 083102.

\bibitem{Watanabe2018Shape}
K.~Watanabe, B.~Jinnai, S.~Fukami, H.~Sato, H.~Ohno,
\newblock \emph{Nature Communications} \textbf{2018}, \emph{9}, 1 663.

\bibitem{Hou20192D}
X.~Hou, H.~Chen, Z.~Zhang, S.~Wang, P.~Zhou,
\newblock \emph{Advanced Electronic Materials} \textbf{2019}, \emph{5}, 9
  1800944.

\bibitem{Zhang2020Memory}
Z.~Zhang, Z.~Wang, T.~Shi, C.~Bi, F.~Rao, Y.~Cai, Q.~Liu, H.~Wu, P.~Zhou,
\newblock \emph{InfoMat} \textbf{2020}, \emph{2}, 2 261.

\bibitem{Liu2018Screening}
H.~Liu, J.~T. Sun, M.~Liu, S.~Meng,
\newblock \emph{Journal of Physical Chemistry Letters} \textbf{2018}, \emph{9},
  23 6709.

\bibitem{Littlejohn2019Large}
A.~J. Littlejohn, Z.~Li, Z.~Lu, X.~Sun, P.~Nawarat, Y.~Wang, Y.~Li, T.~Wang,
  Y.~Chen, L.~Zhang, H.~Li, K.~Kisslinger, S.~Shi, J.~Shi, A.~Raeliarijaona,
  W.~Shi, H.~Terrones, K.~M. Lewis, M.~Washington, T.~M. Lu, G.~C. Wang,
\newblock \emph{ACS Applied Nano Materials} \textbf{2019}, \emph{2}, 6 3684.

\bibitem{Chumak2015Magnon}
A.~V. Chumak, V.~I. Vasyuchka, A.~A. Serga, B.~Hillebrands,
\newblock \emph{Nature Physics} \textbf{2015}, \emph{11}, 6 453.

\bibitem{MacNeill2016Control}
D.~MacNeill, G.~M. Stiehl, M.~H. Guimaraes, R.~A. Buhrman, J.~Park, D.~C.
  Ralph,
\newblock \emph{Nature Physics} \textbf{2016}, \emph{13}, 3 300.

\bibitem{Sadovnikov2017Spin}
A.~V. Sadovnikov, C.~S. Davies, V.~V. Kruglyak, D.~V. Romanenko, S.~V. Grishin,
  E.~N. Beginin, Y.~P. Sharaevskii, S.~A. Nikitov,
\newblock \emph{Physical Review B} \textbf{2017}, \emph{96} 060401.

\bibitem{Zhang2019Van}
W.~Zhang, P.~K.~J. Wong, R.~Zhu, A.~T.~S. Wee,
\newblock \emph{InfoMat} \textbf{2019}, \emph{1}, 4 479.

\bibitem{Burch2018Magnetism}
K.~S. Burch, D.~Mandrus, J.~G. Park,
\newblock \emph{Nature} \textbf{2018}, \emph{563}, 7729 47.

\bibitem{Song2018Giant}
T.~Song, X.~Cai, M.~W.~Y. Tu, X.~Zhang, B.~Huang, N.~P. Wilson, K.~L. Seyler,
  L.~Zhu, T.~Taniguchi, K.~Watanabe, M.~A. McGuire, D.~H. Cobden, D.~Xiao,
  W.~Yao, X.~Xu,
\newblock \emph{Science} \textbf{2018}, \emph{360}, 6394 1214.

\bibitem{Frisenda2018Recent}
R.~Frisenda, E.~Navarro-Moratalla, P.~Gant, D.~{P{\'{e}}rez De Lara},
  P.~Jarillo-Herrero, R.~V. Gorbachev, A.~Castellanos-Gomez,
\newblock \emph{Chemical Society Reviews} \textbf{2018}, \emph{47}, 1 53.

\bibitem{Gibertini2019Magnetic}
M.~Gibertini, M.~Koperski, A.~F. Morpurgo, K.~S. Novoselov,
\newblock \emph{Nature Nanotechnology} \textbf{2019}, \emph{14}, 5 408.

\bibitem{Omer2008Energy}
A.~M. Omer,
\newblock \emph{Renewable and Sustainable Energy Reviews} \textbf{2008},
  \emph{12}, 9 2265.

\bibitem{Shen2009Recent}
B.~Shen, J.~Sun, F.~Hu, H.~Zhang, Z.~Cheng,
\newblock \emph{Advanced Materials} \textbf{2009}, \emph{21}, 45 4545.

\bibitem{Gutfleisch2011Magnetic}
O.~Gutfleisch, M.~A. Willard, E.~Br{\"{u}}ck, C.~H. Chen, S.~G. Sankar, J.~P.
  Liu,
\newblock \emph{Advanced Materials} \textbf{2011}, \emph{23}, 7 821.

\bibitem{Gutfleisch2016Mastering}
O.~Gutfleisch, T.~Gottschall, M.~Fries, D.~Benke, I.~Radulov, K.~P. Skokov,
  H.~Wende, M.~Gruner, M.~Acet, P.~Entel, et~al.,
\newblock \emph{Philosophical Transactions of the Royal Society A:
  Mathematical, Physical and Engineering Sciences} \textbf{2016}, \emph{374},
  2074 20150308.

\bibitem{Li2020Understanding}
F.~Li, M.~Li, X.~Xu, Z.~Yang, H.~Xu, C.~Jia, K.~Li, J.~He, B.~Li, H.~Wang,
\newblock \emph{Nature Communications} \textbf{2020}, \emph{11}, 1 4190.

\bibitem{Liu2018Anisotropic}
Y.~Liu, C.~Petrovic,
\newblock \emph{Physical Review B} \textbf{2018}, \emph{97}, 17 174418.

\bibitem{Yu2019Large}
X.~Yu, X.~Zhang, Q.~Shi, S.~Tian, H.~Lei, K.~Xu, H.~Hosono,
\newblock \emph{Frontiers of Physics} \textbf{2019}, \emph{14}, 4 6.

\bibitem{Liu2020Anisotropic}
Y.~Liu, C.~Petrovic,
\newblock \emph{Physical Review B} \textbf{2020}, \emph{102}, 1 014424.

\bibitem{Mondal2020Magnetic}
S.~Mondal, A.~Midya, M.~M. Patidar, V.~Ganesan, P.~Mandal,
\newblock \emph{Applied Physics Letters} \textbf{2020}, \emph{117}, 9 092405.

\bibitem{liu2020Critical}
Y.~Liu, M.~Abeykoon, C.~Petrovic, et~al.,
\newblock \emph{Physical Review Research} \textbf{2020}, \emph{2}, 1 013013.

\bibitem{Tran2022Insight}
H.~B. Tran, H.~Momida, Y.~ichiro Matsushita, K.~Shirai, T.~Oguchi,
\newblock \emph{Acta Materialia} \textbf{2022}, \emph{231} 117851.

\bibitem{Franco2018Magnetocaloric}
V.~Franco, J.~Bl{\'a}zquez, J.~Ipus, J.~Law, L.~Moreno-Ram{\'\i}rez, A.~Conde,
\newblock \emph{Progress in Materials Science} \textbf{2018}, \emph{93} 112.

\bibitem{Teichert2015Structure}
N.~Teichert, D.~Kucza, O.~Yildirim, E.~Yuzuak, I.~Dincer, A.~Behler, B.~Weise,
  L.~Helmich, A.~Boehnke, S.~Klimova, A.~Waske, Y.~Elerman, A.~H\"utten,
\newblock \emph{Physical Review B} \textbf{2015}, \emph{91} 184405.

\bibitem{Dontgen2015Temp}
J.~D{\"{o}}ntgen, J.~Rudolph, T.~Gottschall, O.~Gutfleisch, S.~Salomon,
  A.~Ludwig, D.~H{\"{a}}gele,
\newblock \emph{Applied Physics Letters} \textbf{2015}, \emph{106}, 3 032408.

\bibitem{Nguyen2021Magnetocaloric}
D.~Nguyen~Ba, Y.~Zheng, L.~Becerra, M.~Marangolo, M.~Almanza, M.~LoBue,
\newblock \emph{Physical Review Applied} \textbf{2021}, \emph{15} 064045.

\bibitem{Zhang2015Robust}
W.-B. Zhang, Q.~Qu, P.~Zhu, C.-H. Lam,
\newblock \emph{Journal of Materials Chemistry C} \textbf{2015}, \emph{3}
  12457.

\bibitem{Xu2021A}
C.~Xu, J.~Zhang, Z.~Guo, S.~Zhang, X.~Yuan, L.~Wang,
\newblock \emph{Journal of Physics: Condensed Matter} \textbf{2021}, \emph{33},
  19 195804.

\bibitem{Guo2018Chromium}
Y.~Guo, Y.~Zhang, S.~Yuan, B.~Wang, J.~Wang,
\newblock \emph{Nanoscale} \textbf{2018}, \emph{10}, 37 18036.

\bibitem{Han2020Prediction}
R.~Han, Z.~Jiang, Y.~Yan,
\newblock \emph{Journal of Physical Chemistry C} \textbf{2020}, \emph{124}, 14
  7956.

\bibitem{Zhang2019Direct}
Z.~Zhang, J.~Shang, C.~Jiang, A.~Rasmita, W.~Gao, T.~Yu,
\newblock \emph{Nano Letters} \textbf{2019}, \emph{19}, 5 3138.

\bibitem{Zhang2019Magnetism}
T.~Zhang, Y.~Wang, H.~Li, F.~Zhong, J.~Shi, M.~Wu, Z.~Sun, W.~Shen, B.~Wei,
  W.~Hu, X.~Liu, L.~Huang, C.~Hu, Z.~Wang, C.~Jiang, S.~Yang, Q.~M. Zhang,
  Z.~Qu,
\newblock \emph{ACS Nano} \textbf{2019}, \emph{13}, 10 11353.

\bibitem{Lee2021Magnetic}
K.~Lee, A.~H. Dismukes, E.~J. Telford, R.~A. Wiscons, J.~Wang, X.~Xu,
  C.~Nuckolls, C.~R. Dean, X.~Roy, X.~Zhu,
\newblock \emph{Nano Letters} \textbf{2021}, \emph{21}, 8 3511.

\bibitem{Liu2016Exfoliating}
J.~Liu, Q.~Sun, Y.~Kawazoe, P.~Jena,
\newblock \emph{Physical Chemistry Chemical Physics} \textbf{2016}, \emph{18}
  8777.

\bibitem{Rong2020Ferromagnetism}
Q.-Y. Rong, A.-M. Hu, X.-H. Zhang, L.-L. Wang, W.-Z. Xiao,
\newblock \emph{Journal of Magnetism and Magnetic Materials} \textbf{2020},
  \emph{515} 167310.

\bibitem{Xiao2020Modulating}
J.~Xiao, D.~Legut, W.~Luo, H.~Guo, X.~Liu, R.~Zhang, Q.~Zhang,
\newblock \emph{Physical Review B} \textbf{2020}, \emph{101}, 1 014431.

\bibitem{Xiao2018Theoretical}
T.~Xiao, G.~Wang, Y.~Liao,
\newblock \emph{Chemical Physics} \textbf{2018}, \emph{513}, April 182.

\bibitem{Zhang2019Super}
F.~Zhang, Y.~C. Kong, R.~Pang, L.~Hu, P.~L. Gong, X.~Q. Shi, Z.~K. Tang,
\newblock \emph{New Journal of Physics} \textbf{2019}, \emph{21}, 5.

\bibitem{Jiang2021Recent}
X.~Jiang, Q.~Liu, J.~Xing, N.~Liu, Y.~Guo, Z.~Liu, J.~Zhao,
\newblock \emph{Applied Physics Reviews} \textbf{2021}, \emph{8}, 3 031305.

\bibitem{Sadhukhan2022Spin}
B.~Sadhukhan, A.~Bergman, Y.~O. Kvashnin, J.~Hellsvik, A.~Delin,
\newblock \emph{Physical Review B} \textbf{2022}, \emph{105}, 10 104418.

\bibitem{Staros2022A}
D.~Staros, G.~Hu, J.~Tiihonen, R.~Nanguneri, J.~Krogel, M.~C. Bennett,
  O.~Heinonen, P.~Ganesh, B.~Rubenstein,
\newblock \emph{The Journal of Chemical Physics} \textbf{2022}, \emph{156}, 1
  014707.

\bibitem{Midya2011Anisotropic}
A.~Midya, S.~N. Das, P.~Mandal, S.~Pandya, V.~Ganesan,
\newblock \emph{Physical Review B} \textbf{2011}, \emph{84} 235127.

\bibitem{Zhang2019Review}
Y.~Zhang,
\newblock \emph{Journal of Alloys and Compounds} \textbf{2019}, \emph{787}
  1173.

\bibitem{Phan2004Large}
M.-H. Phan, S.-C. Yu, N.~H. Hur, Y.-H. Jeong,
\newblock \emph{Journal of Applied Physics} \textbf{2004}, \emph{96}, 2 1154.

\bibitem{Patra2009Multifunctionality}
M.~Patra, K.~De, S.~Majumdar, S.~Giri,
\newblock \emph{Applied Physics Letters} \textbf{2009}, \emph{94} 092506.

\bibitem{Aliev2018Magnetic}
A.~M. Aliev, A.~B. Batdalov, L.~N. Khanov,
\newblock \emph{Applied Physics Letters} \textbf{2018}, \emph{112}, 14 142407.

\bibitem{Gamzatov2018Correlation}
A.~G. Gamzatov, A.~M. Aliev, P.~D. Yen, L.~Khanov, K.~X. Hau, T.~D. Thanh,
  N.~T. Dung, S.~C. Yu,
\newblock \emph{Journal of Applied Physics} \textbf{2018}, \emph{124}, 18
  183902.

\bibitem{Ram2018Review}
N.~R. Ram, M.~Prakash, U.~Naresh, N.~S. Kumar, T.~S. Sarmash, T.~Subbarao,
  R.~J. Kumar, G.~R. Kumar, K.~C.~B. Naidu,
\newblock \emph{Journal of Superconductivity and Novel Magnetism}
  \textbf{2018}, \emph{31}, 7 1971.

\bibitem{Jia2009Magnetocaloric}
L.~Jia, J.~R. Sun, J.~Shen, Q.~Y. Dong, J.~D. Zou, B.~Gao, T.~Y. Zhao, H.~W.
  Zhang, F.~X. Hu, B.~G. Shen,
\newblock \emph{Journal of Applied Physics} \textbf{2009}, \emph{105}, 7
  07A924.

\bibitem{Gottschall2019Making}
T.~Gottschall, K.~P. Skokov, M.~Fries, A.~Taubel, I.~Radulov, F.~Scheibel,
  D.~Benke, S.~Riegg, O.~Gutfleisch,
\newblock \emph{Advanced Energy Materials} \textbf{2019}, \emph{9}, 34 1970130.

\bibitem{Castro2020Machine}
P.~B. de~Castro, K.~Terashima, T.~D. Yamamoto, Z.~Hou, S.~Iwasaki,
  R.~Matsumoto, S.~Adachi, Y.~Saito, P.~Song, H.~Takeya, Y.~Takano,
\newblock \emph{NPG Asia Materials} \textbf{2020}, \emph{12}, 1 1.

\bibitem{Zarkevich2020Viable}
N.~A. Zarkevich, V.~I. Zverev,
\newblock \emph{Crystals} \textbf{2020}, \emph{10}, 9 815.

\bibitem{Cantos2021Layer}
F.~Cantos-Prieto, A.~Falin, M.~Alliati, D.~Qian, R.~Zhang, T.~Tao, M.~R.
  Barnett, E.~J.~G. Santos, L.~H. Li, E.~Navarro-Moratalla,
\newblock \emph{Nano Letters} \textbf{2021}, \emph{21}, 8 3379.

\bibitem{Kresse1996Efficient}
G.~Kresse, J.~Furthm\"uller,
\newblock \emph{Physical Review B} \textbf{1996}, \emph{54} 11169.

\bibitem{Kresse1999From}
G.~Kresse, D.~Joubert,
\newblock \emph{Physical Review B} \textbf{1999}, \emph{59} 1758.

\bibitem{Blochl1994Projector}
P.~E. Bl\"ochl,
\newblock \emph{Physical Review B} \textbf{1994}, \emph{50} 17953.

\bibitem{Kresse1996Efficiency}
G.~Kresse, J.~Furthmüller,
\newblock \emph{Computational Materials Science} \textbf{1996}, \emph{6}, 1 15.

\bibitem{Perdew1996Generalized}
J.~P. Perdew, K.~Burke, M.~Ernzerhof,
\newblock \emph{Physical Review Letters} \textbf{1996}, \emph{77} 3865.

\bibitem{Dillon1965Magnetization}
J.~F. Dillon, C.~E. Olson,
\newblock \emph{Journal of Applied Physics} \textbf{1965}, \emph{36}, 3 1259.

\bibitem{LopezPaz2022Dynamic}
S.~A. L{\'{o}}pez-Paz, Z.~Guguchia, V.~Y. Pomjakushin, C.~Witteveen,
  A.~Cervellino, H.~Luetkens, N.~Casati, A.~F. Morpurgo, F.~O. von Rohr,
\newblock \emph{Nature Communications} \textbf{2022}, \emph{13}, 1 1.

\bibitem{Wilson2021Interlayer}
N.~P. Wilson, K.~Lee, J.~Cenker, K.~Xie, A.~H. Dismukes, E.~J. Telford,
  J.~Fonseca, S.~Sivakumar, C.~Dean, T.~Cao, X.~Roy, X.~Xu, X.~Zhu,
\newblock \emph{Nature Materials} \textbf{2021}, \emph{20}, 12 1657.

\bibitem{Avsar2022Highly}
A.~Avsar,
\newblock \emph{Nature Materials} \textbf{2022}, \emph{21}, 7 731.

\bibitem{Monkhorst1976Special}
H.~J. Monkhorst, J.~D. Pack,
\newblock \emph{Physical Review B} \textbf{1976}, \emph{13} 5188.

\bibitem{Togo2015First}
A.~Togo, I.~Tanaka,
\newblock \emph{Scripta Materialia} \textbf{2015}, \emph{108} 1359.

\bibitem{Chaput2011Phonon}
L.~Chaput, A.~Togo, I.~Tanaka, G.~Hug,
\newblock \emph{Physical Review B} \textbf{2011}, \emph{84} 094302.

\bibitem{Togo2010First}
A.~Togo, L.~Chaput, I.~Tanaka, G.~Hug,
\newblock \emph{Physical Review B} \textbf{2010}, \emph{81} 174301.

\bibitem{Skubic2008A}
B.~Skubic, J.~Hellsvik, L.~Nordström, O.~Eriksson,
\newblock \emph{Journal of Physics: Condensed Matter} \textbf{2008}, \emph{20},
  31 315203.

\bibitem{Evans2014Atomistic}
R.~F. Evans, W.~J. Fan, P.~Chureemart, T.~A. Ostler, M.~O. Ellis, R.~W.
  Chantrell,
\newblock \emph{Journal of Physics: Condensed Matter} \textbf{2014}, \emph{26},
  10 103202.

\bibitem{Pecharsky1999Magnetocaloric}
V.~K. Pecharsky, K.~A. Gschneidner,
\newblock \emph{Journal of Applied Physics} \textbf{1999}, \emph{86}, 1 565.

\bibitem{Gschneidner2000Magnetocaloric}
K.~A. Gschneidner, V.~K. Pecharsky,
\newblock \emph{Annual Review of Materials Science} \textbf{2000}, \emph{30}, 1
  387.

\end{thebibliography}

\begin{figure}[!h]
\textbf{Table of Contents}\\
\centering
\medskip
 \includegraphics[width=18cm]{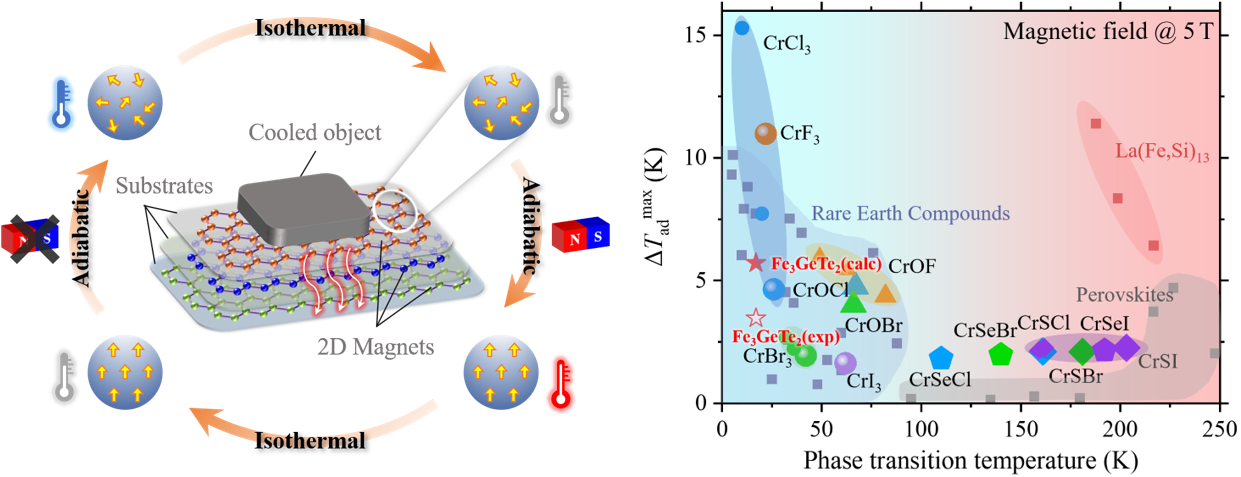}
  \medskip
  \caption*{The existence of giant magnetocaloric effect (MCE) and its strain tunability in monolayer magnets such as CrX$_3$ (X\,=\,F,~Cl,~Br,~I), CrAX (A\,=\,O,~S,~Se;~X\,=\,F,~Cl,~Br,~I), and Fe$_3$GeTe$_2$ are revealed through multiscale calculations. MCE of 2D magnets is theoretically comparable to that of bulk materials and can be remarkably tuned by strain. In particular, CrF$_3$ possesses the state-of-the-art MCE at low temperatures. These findings advocate the giant-MCE monolayer magnets, opening new opportunities for magnetic cooling at nanoscale.}
\end{figure}

\end{sloppypar}
\end{spacing}



\section*{Supplementary material}

# Supplementary information

# Giant magnetocaloric effect in magnets down to the monolayer limit

Weiwei He,[†] Yan Yin,[†] Qihua Gong,[*,†] Richard F. L. Evans,[‡] Oliver Gutfleisch,[¶] Baixiang Xu,[¶] Min Yi,[*,†] and Wanlin Guo[*,†]

[†]State Key Lab of Mechanics and Control of Mechanical Structures & Key Lab for Intelligent Nano Materials and Devices of Ministry of Education & Institute for Frontier Science, Nanjing University of Aeronautics and Astronautics (NUAA), Nanjing 210016, China

[‡]Department of Physics, The University of York, York YO105DD, United Kingdom

[¶]Institute of Materials Science, Technische Universität Darmstadt, Darmstadt 64287, Germany

E-mail: gongqihua@nuaa.edu.cn; yimin@nuaa.edu.cn; wlguo@nuaa.edu.cn

1

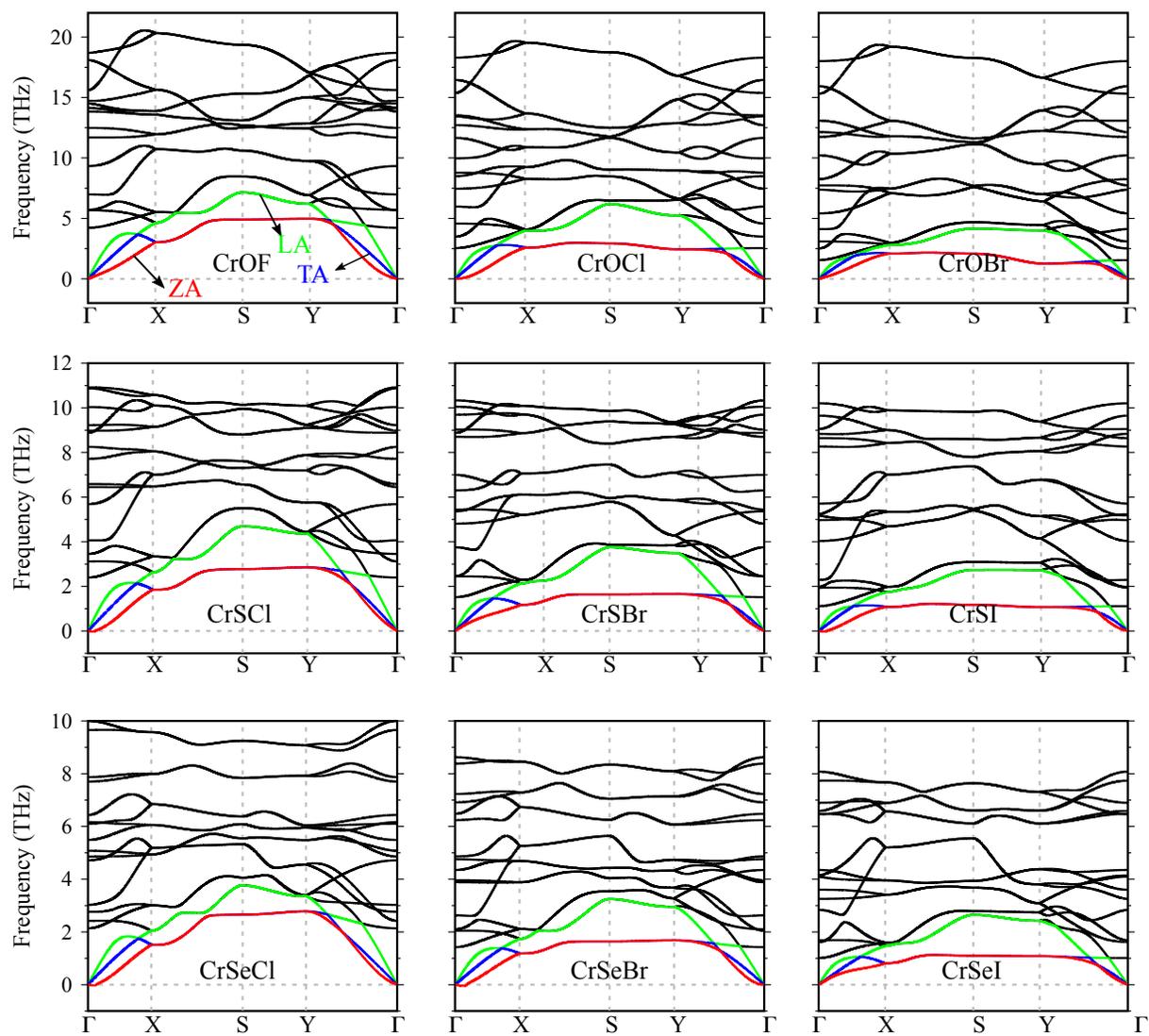

Fig. S1. Phonon dispersion spectra of 2D CrAX.

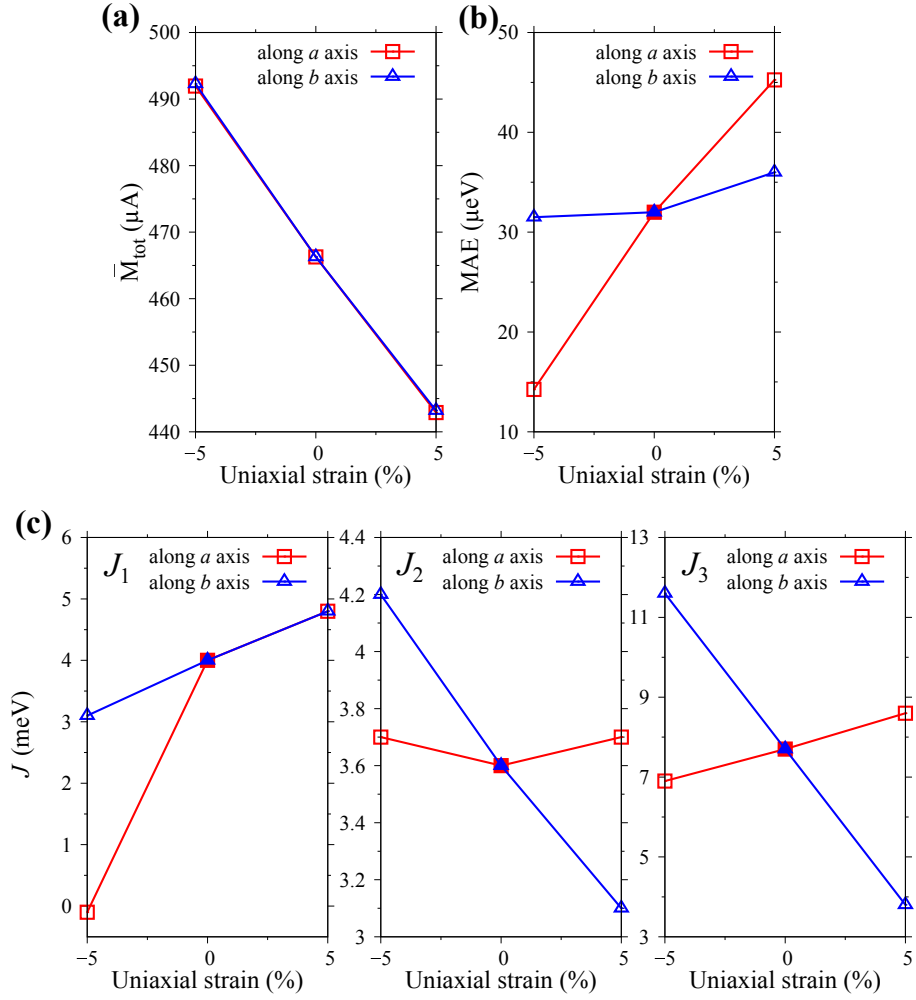

Fig. S2. Strain-tunable magnetic properties of CrOF: (a) net magnetic moment per unit area, (b) MAE, and (c) exchange parameters $J_1$, $J_2$ and $J_3$.

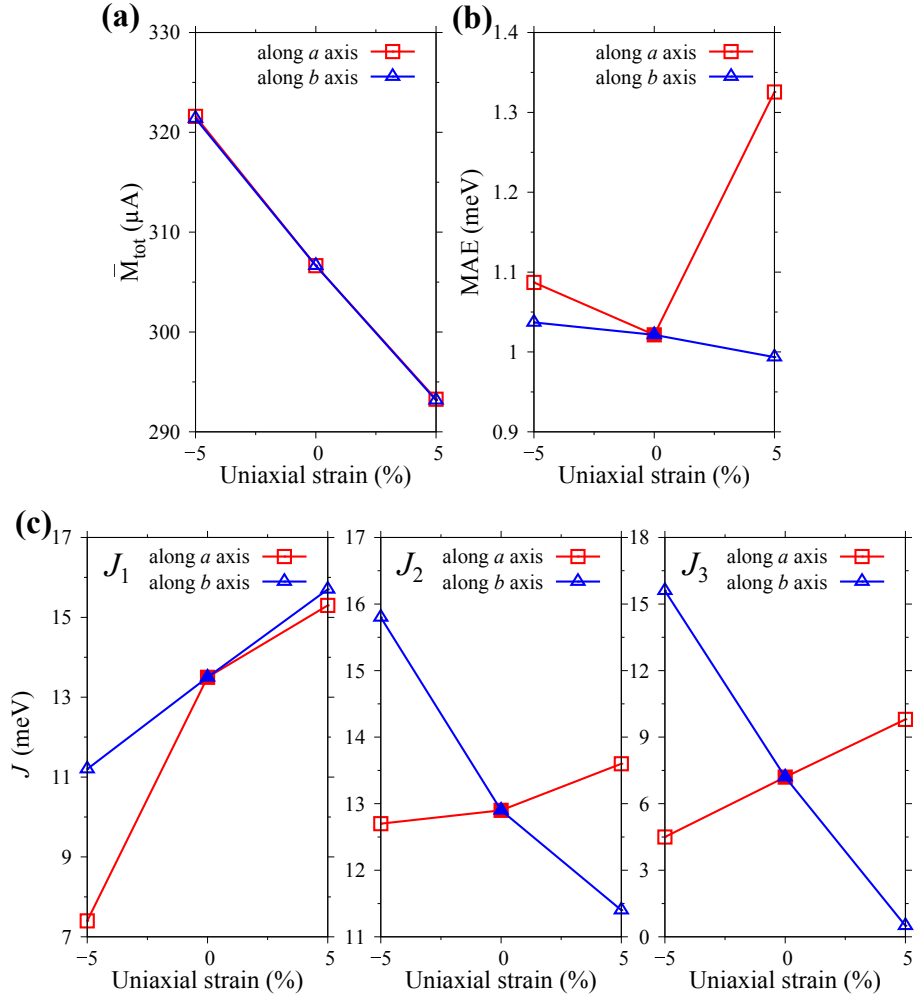

Fig. S3. Strain-tunable magnetic properties of CrSI: (a) net magnetic moment per unit area, (b) MAE, and (c) exchange parameters $J_1$, $J_2$ and $J_3$.

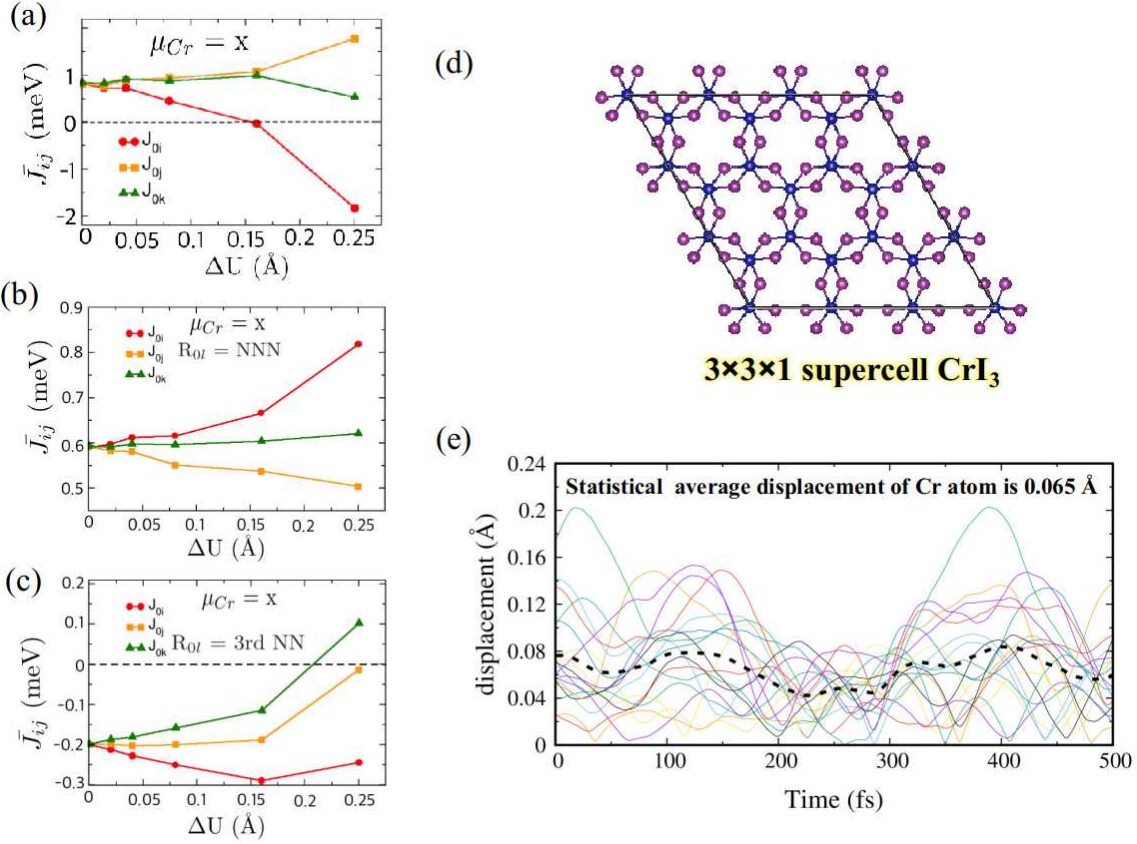

Fig. S4. Temperature effect in exchange interaction. The effect of displacement on (a) the nearest-neighbor (NN), (b) next NN, and (c) third NN exchange interaction of monolayer $CrI_3$ (figures from ref[1]). (d) The calculated structure of the 3×3×1 supercell $CrI_3$ and (e)the calculated displacement of each time step at 60 K by *ab-initio* molecular dynamics simulations. The average displacement of Cr is estimated as 0.065 Å. As previously discussed,[1,2] exchange parameters could be affected by finite temperature induced atom displacements and are found only slightly changed when the Cr displacement is less than 0.1 Å.[1] However, the critical temperature of $CrI_3$ is around 60 K, at which Cr atoms are averagely displaced by about 0.065 Å (d, e). This small displacement will not notably affect the magnetic properties, indicating the reasonable approximation of applying zero-K magnetic parameters in the classic spin Hamiltonians. Thus, the finite-temperature effect on MCE at low temperatures should be inconspicuous. Nevertheless, further in-depth studies are required to calculate the magnetic properties by considering the atom displacements from the vibration modes of 2D magnets if elevated temperatures are of interests.

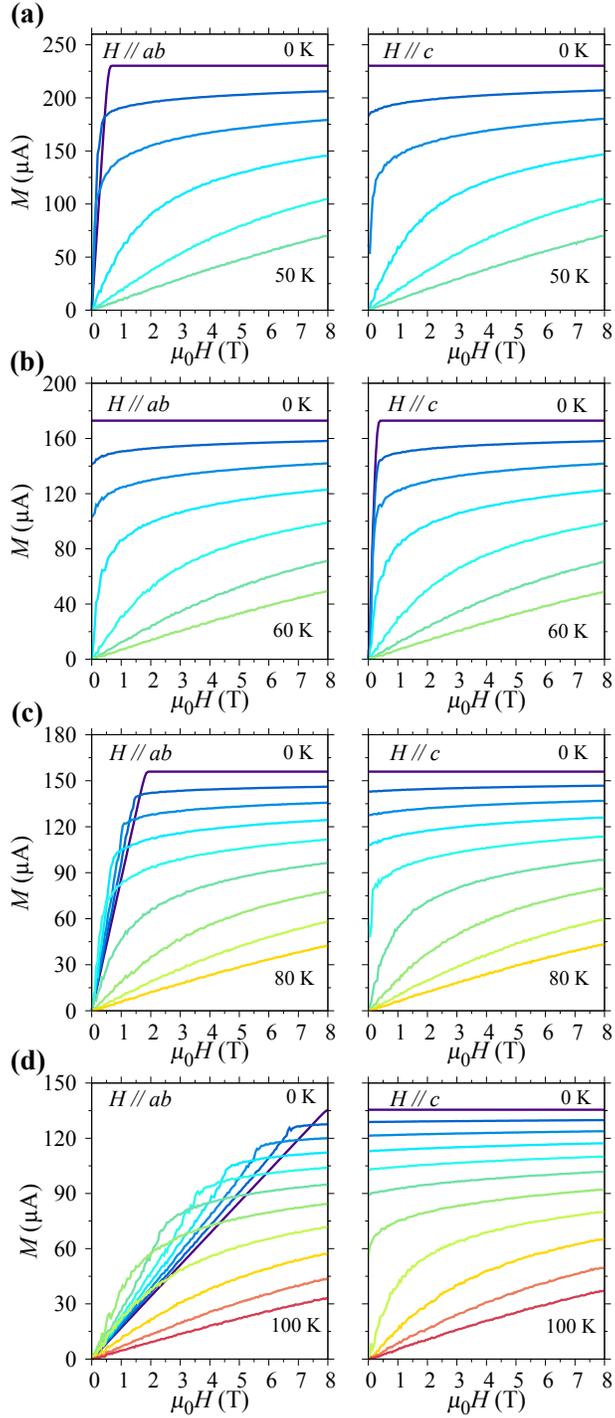

Fig. S5. Isothermal magnetization curves with field up to 8 T applied in the $ab$ plane (left) and along the $c$ axis (right) for (a) $CrF_3$, (b) $CrCl_3$, (c) $CrBr_3$, and (d) $CrI_3$. The curves are displayed every 10 K.

6

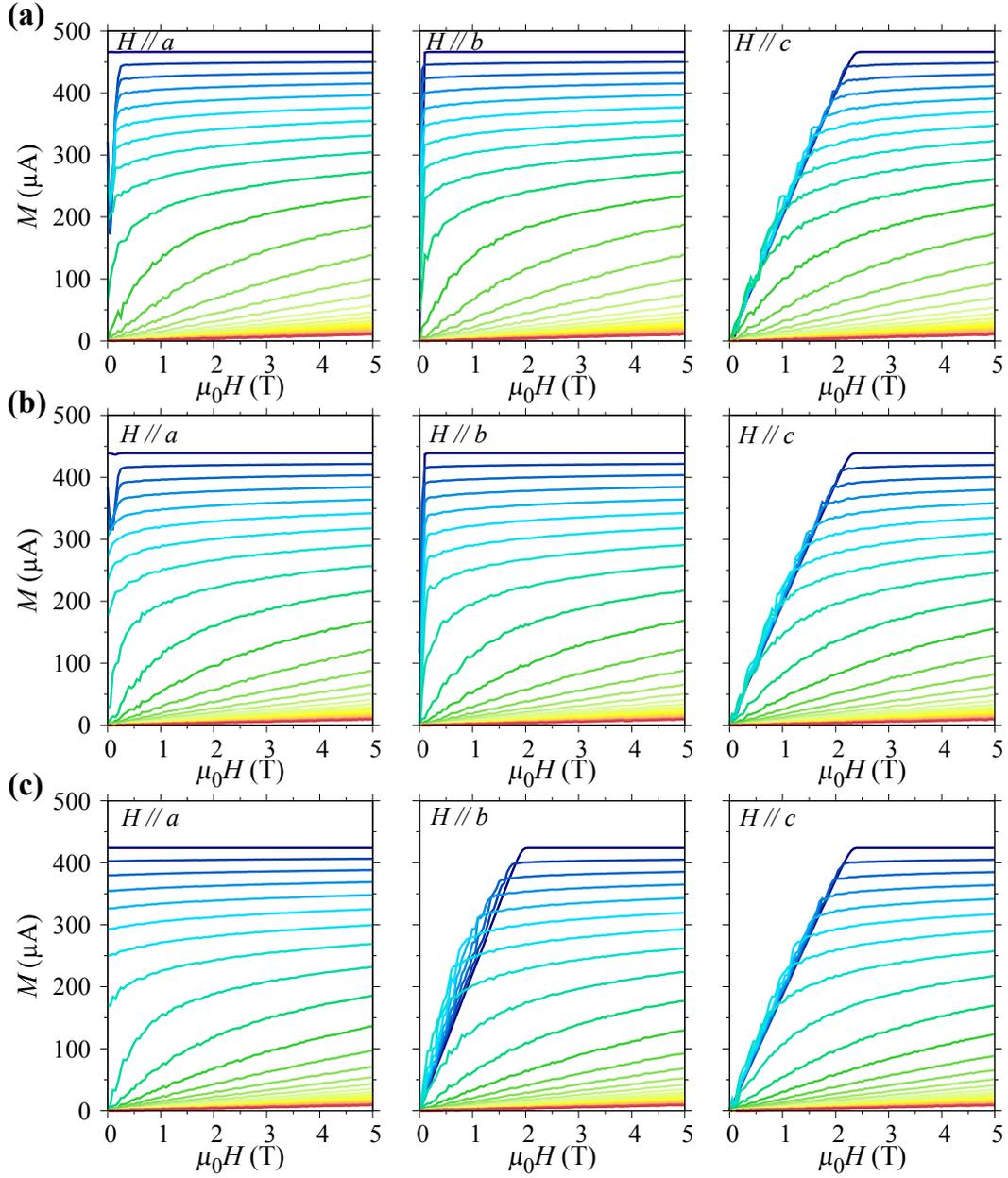

Fig. S6. Isothermal magnetization curves with field up to 5 T applied along the $a$ axis (left), $b$ axis (middle), and $c$ axis (right) for (a) CrOF, (b) CrOCl, and (c) CrOBr. The curves are displayed every 10 K with a temperature range from 0 K to 300 K.

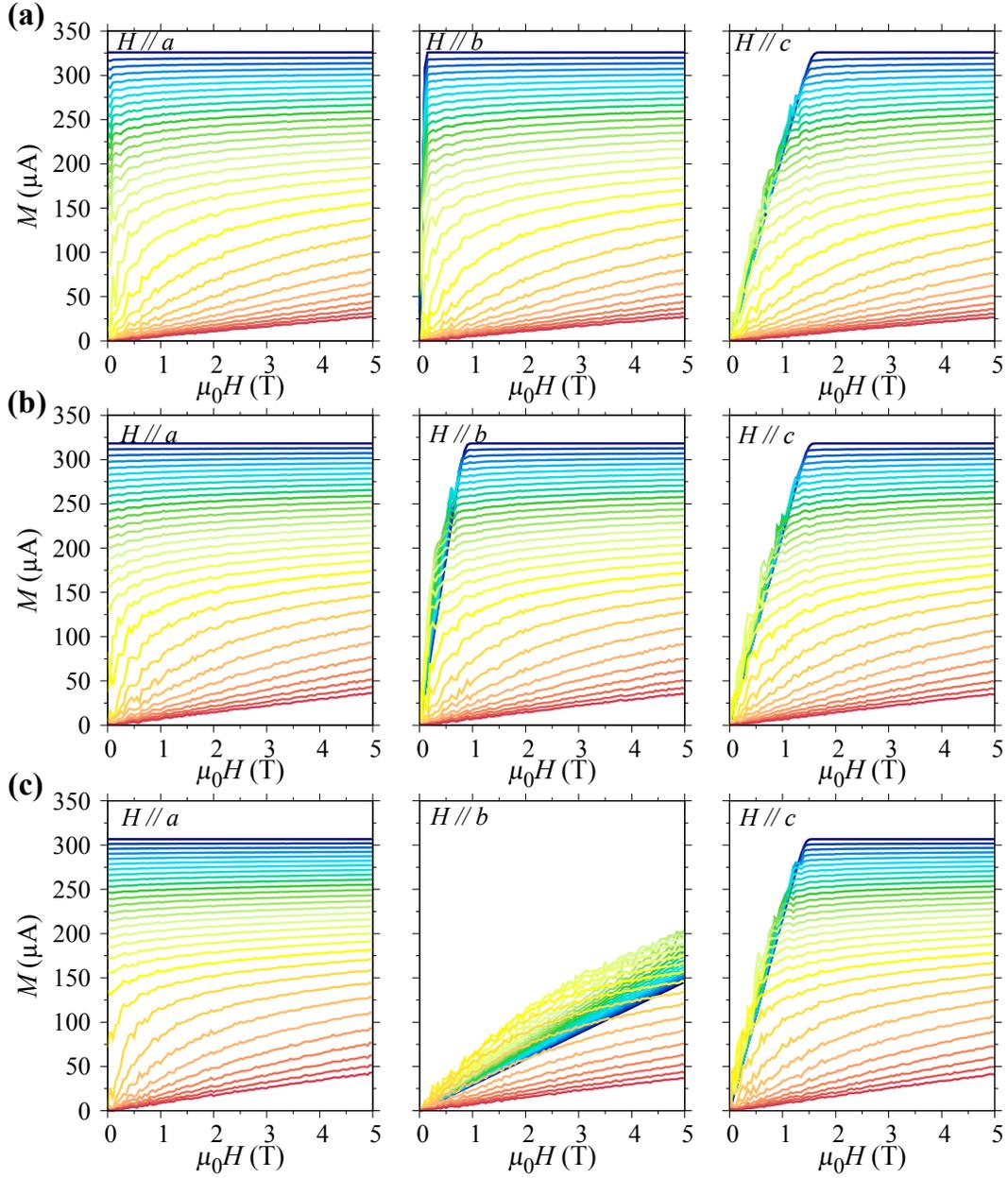

Fig. S7. Isothermal magnetization curves with field up to 5 T applied along the $a$ axis (left), $b$ axis (middle) and $c$ axis (right) for (a) CrSCl, (b) CrSBr, and (c) CrSI. The curves are displayed every 10 K with a temperature range from 0 K to 300 K.

8

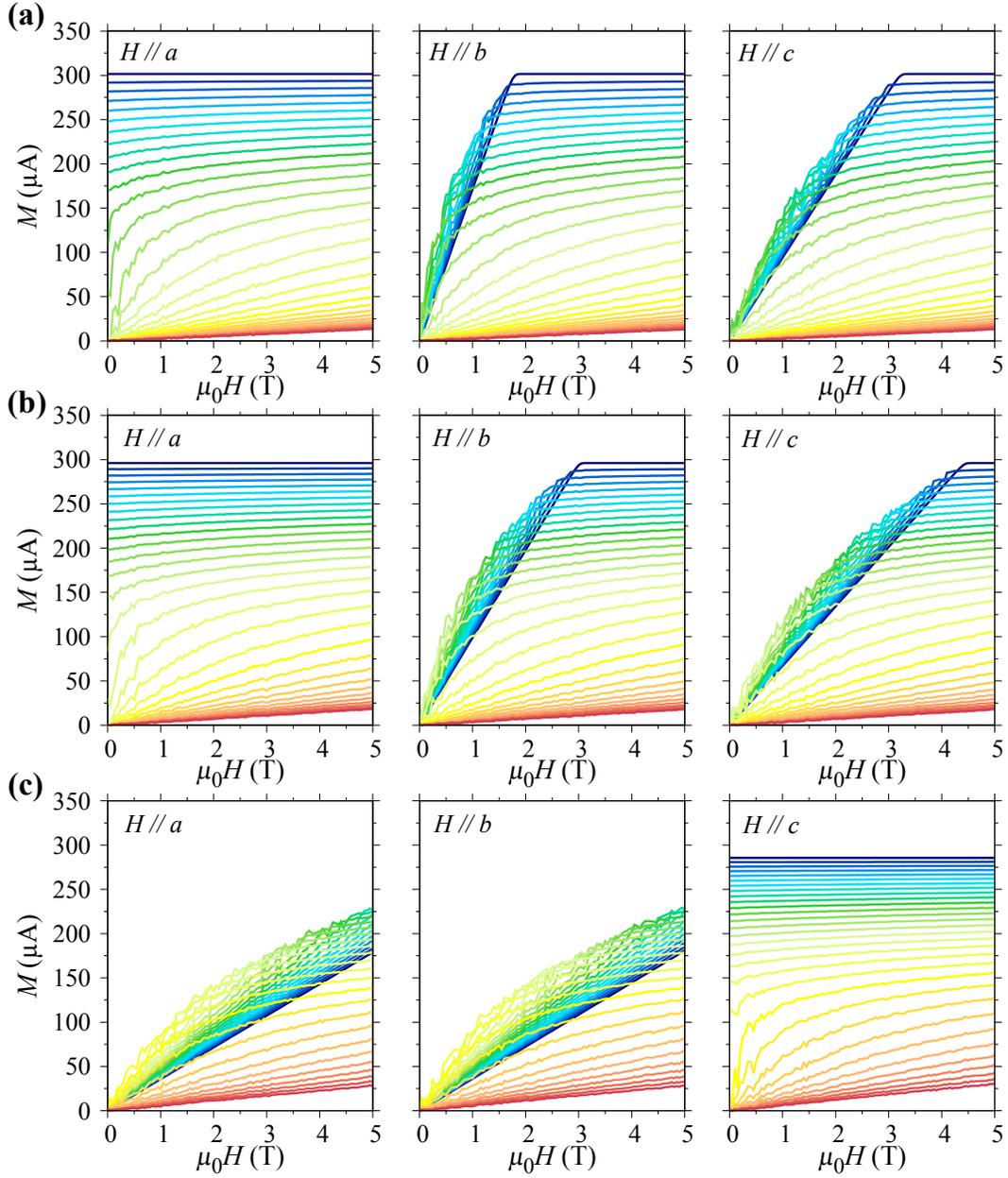

Fig. S8. Isothermal magnetization curves with field up to 5 T applied along the $a$ axis (left), $b$ axis (middle) and $c$ axis (right) for (a) CrSeCl, (b) CrSeBr, and (c) CrSeI. The curves are displayed every 10 K with a temperature range from 0 K to 300 K.

9

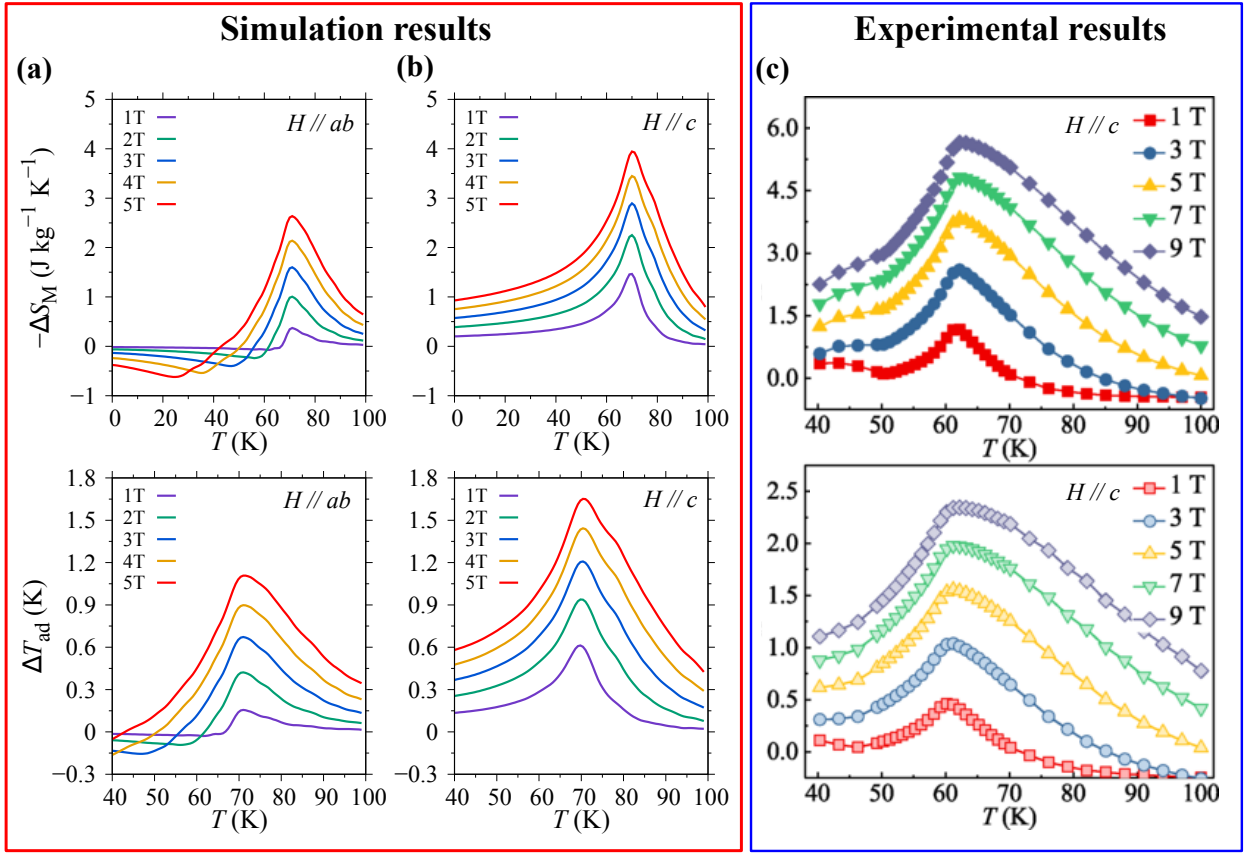

Fig. S9. Simulation and experimental results for MCE of bulk CrI$_3$. Simulated temperature-dependent $-\Delta S_\text{M}$ (top) and $\Delta T_\text{ad}$ (bottom) for (a) $H \parallel ab$ and (b) $H \parallel c$ in different magnetic fields, respectively. Experimental temperature-dependent $-\Delta S_\text{M}$ (top) and $\Delta T_\text{ad}$ (bottom) for (c) $H \parallel c$ in different magnetic fields. Experimental curves in (c) are from ref.[3]

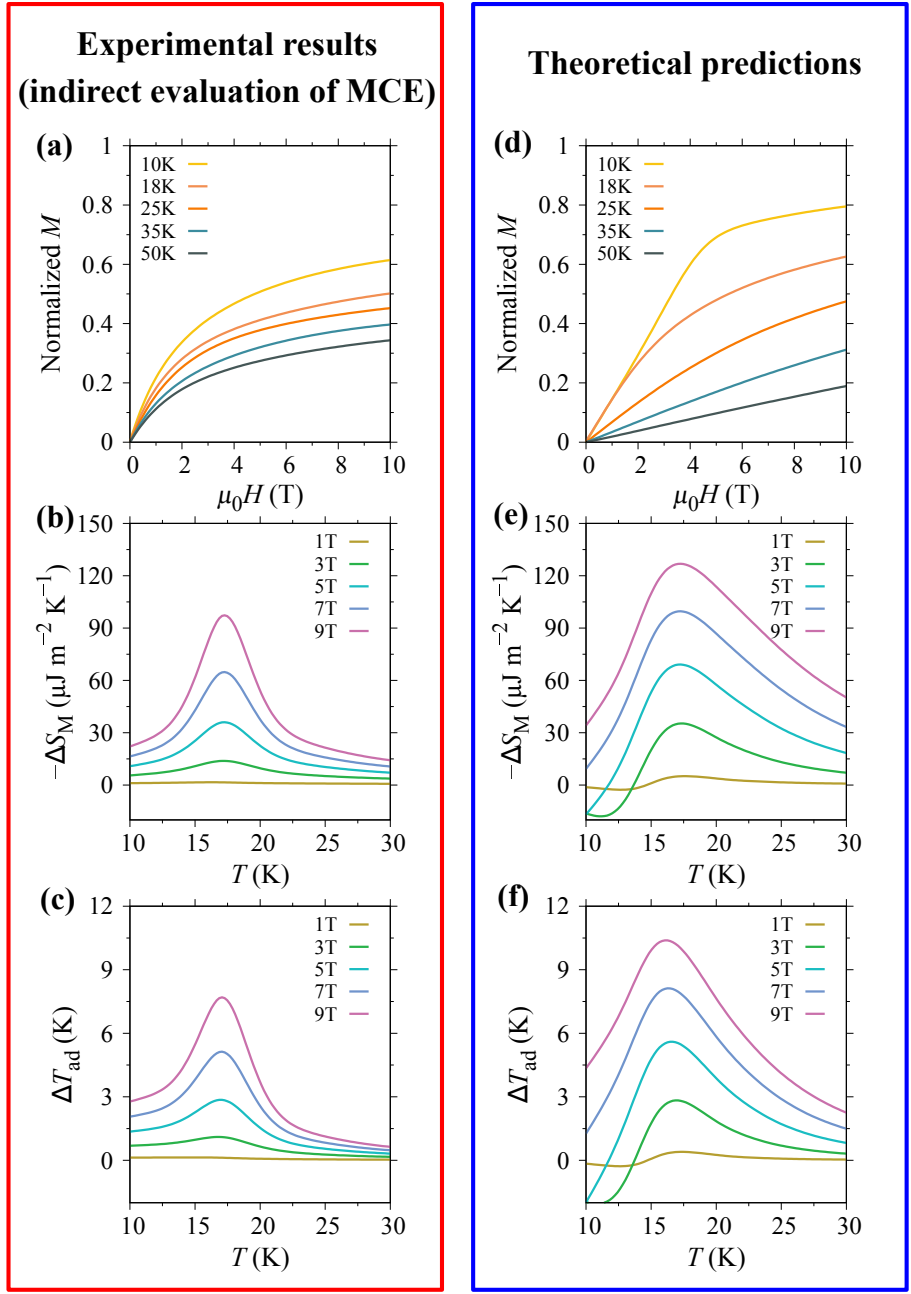

Fig. S10. MCE of monolayer $Fe_3GeTe_2$ from experimental results and theoretical predictions. Temperature-dependent M-H curves: (a) experiment; (d) prediction. $-\Delta S_M$ curves under different magnetic fields: (b) experiment; (e) prediction. $\Delta T_{ad}$ curves under different magnetic fields: (c) experiment; (f) prediction. The experimental data of monolayer $Fe_3GeTe_2$ are from ref.[4]

11

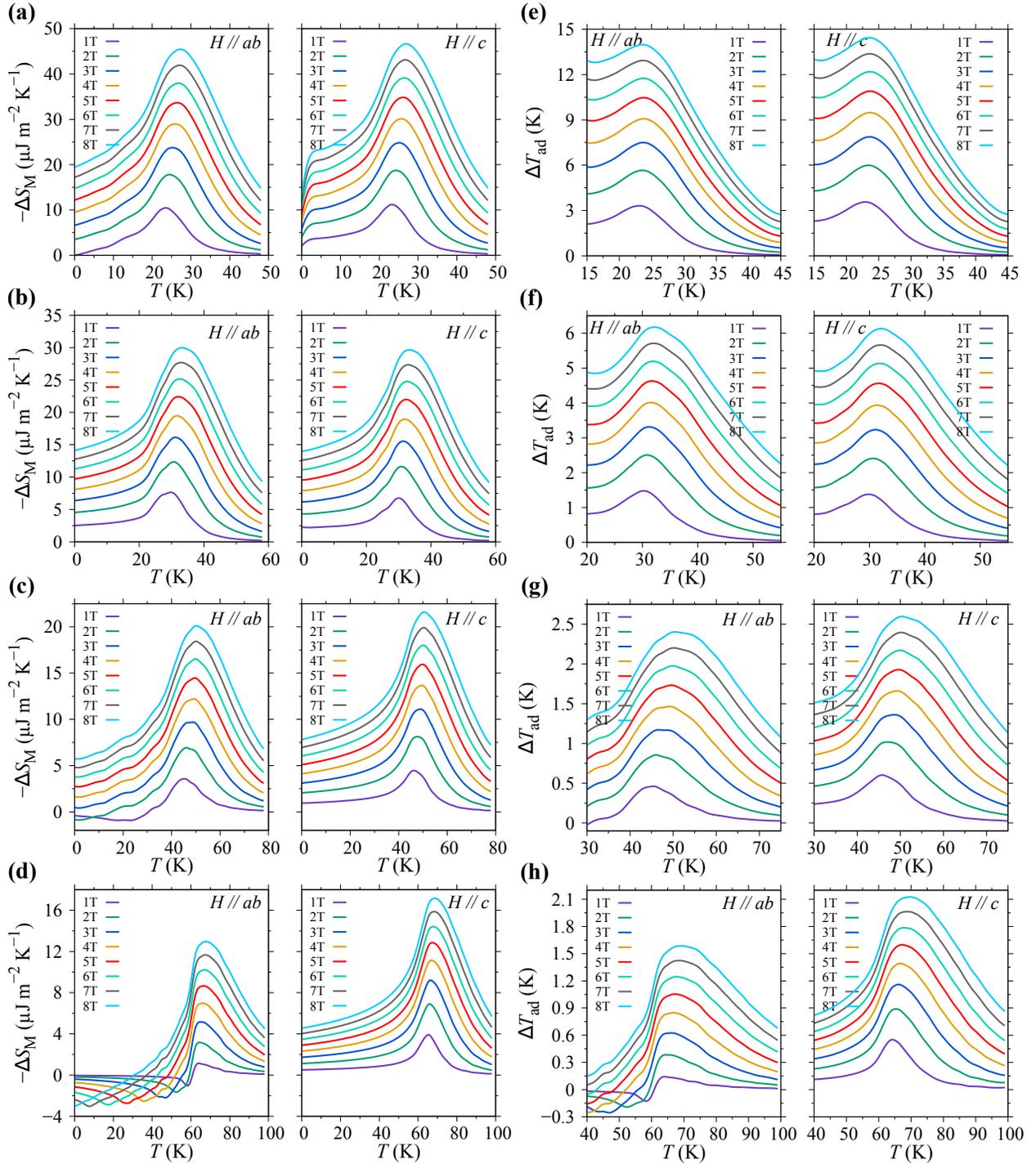

Fig. S11. Temperature dependences of (a-d) $-\Delta S_\mathrm{M}$ and (e-h) $\Delta T_\mathrm{ad}$ of monolayer CrX$_3$ for $H \parallel ab$ and $H \parallel c$ in different magnetic fields, respectively. (a) and (e) are for CrF$_3$, (b) and (f) are for CrCl$_3$, (c) and (g) are for CrBr$_3$, (d) and (h) are for CrI$_3$.

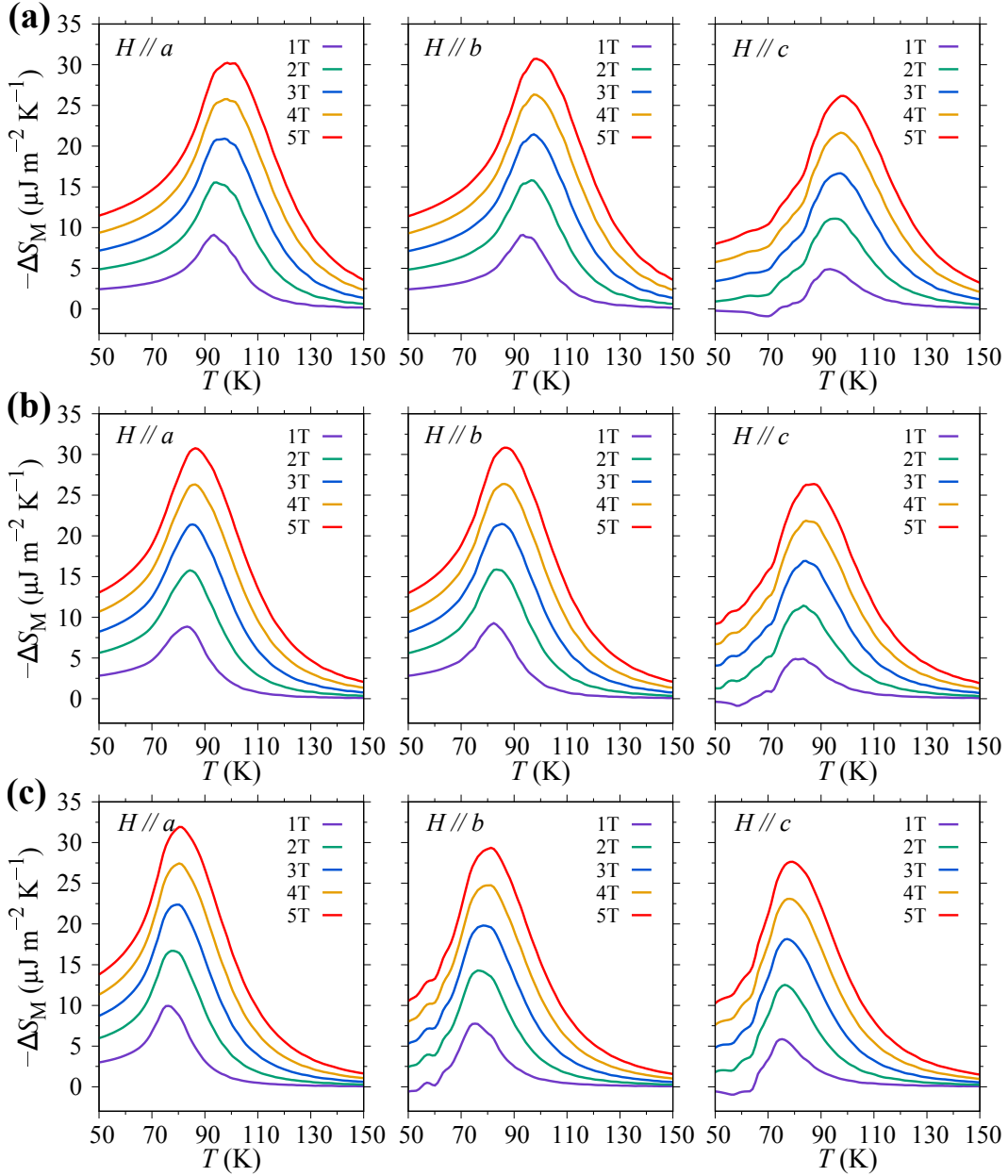

Fig. S12. Temperature dependences of $-\Delta S_{\mathrm{M}}$ of monolayer (a) CrOF, (b) CrOCl and (c) CrOBr for $H \parallel a$, $H \parallel b$, and $H \parallel c$ in different magnetic fields, respectively.

13

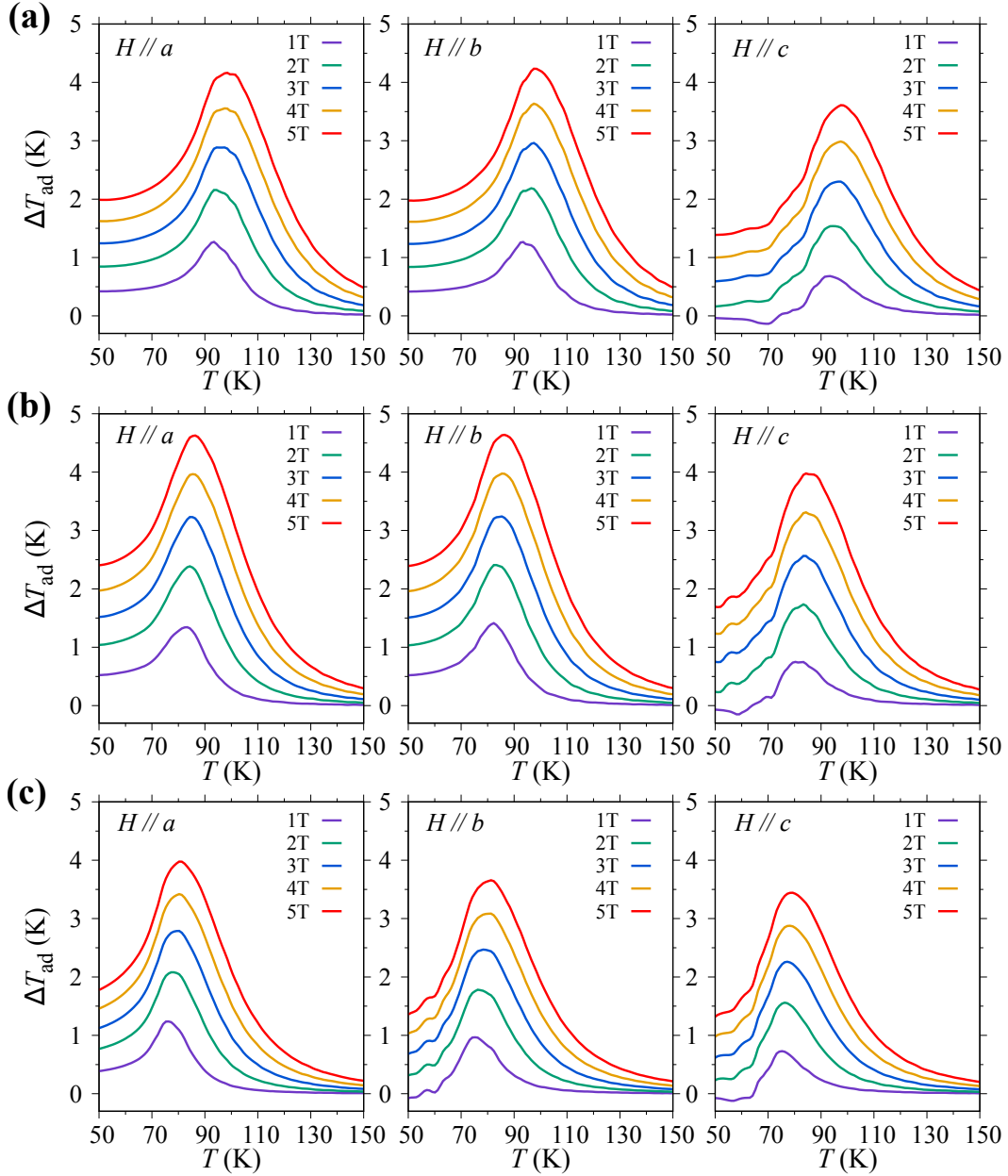

Fig. S13. Temperature dependences of $\Delta T_\mathrm{ad}$ of monolayer (a) CrOF, (b) CrOCl and (c) CrOBr for $H \parallel a$, $H \parallel b$, and $H \parallel c$ in different magnetic fields, respectively.

14

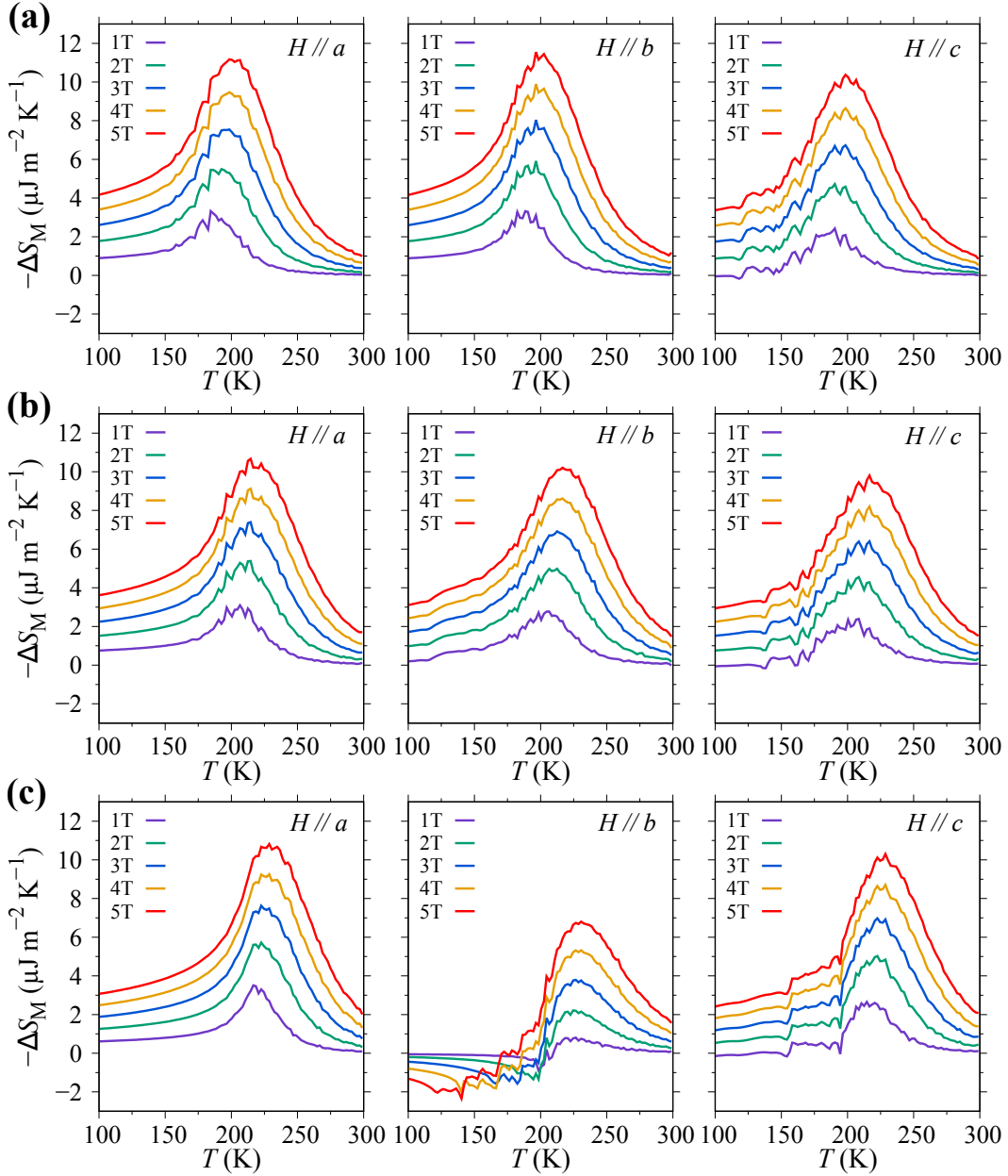

Fig. S14. Temperature dependences of $-\Delta S_\mathrm{M}$ of monolayer (a) CrSCl, (b) CrSBr and (c) CrSI for $H \parallel a$, $H \parallel b$, and $H \parallel c$ in different magnetic fields, respectively.

15

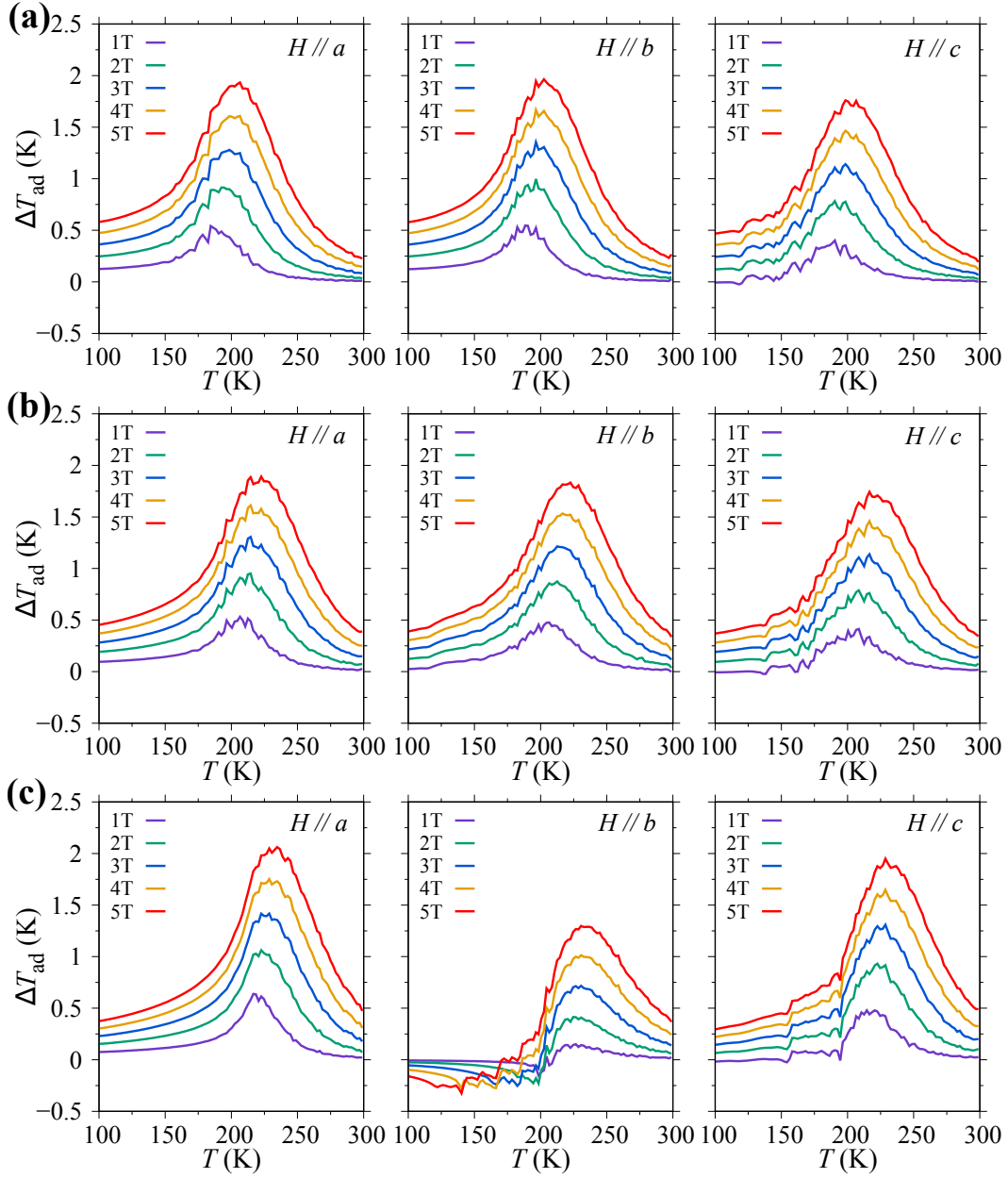

Fig. S15. Temperature dependences of $\Delta T_{\mathrm{ad}}$ of monolayer (a) CrSCl, (b) CrSBr and (c) CrSI for $H \parallel a$, $H \parallel b$, and $H \parallel c$ in different magnetic fields, respectively.

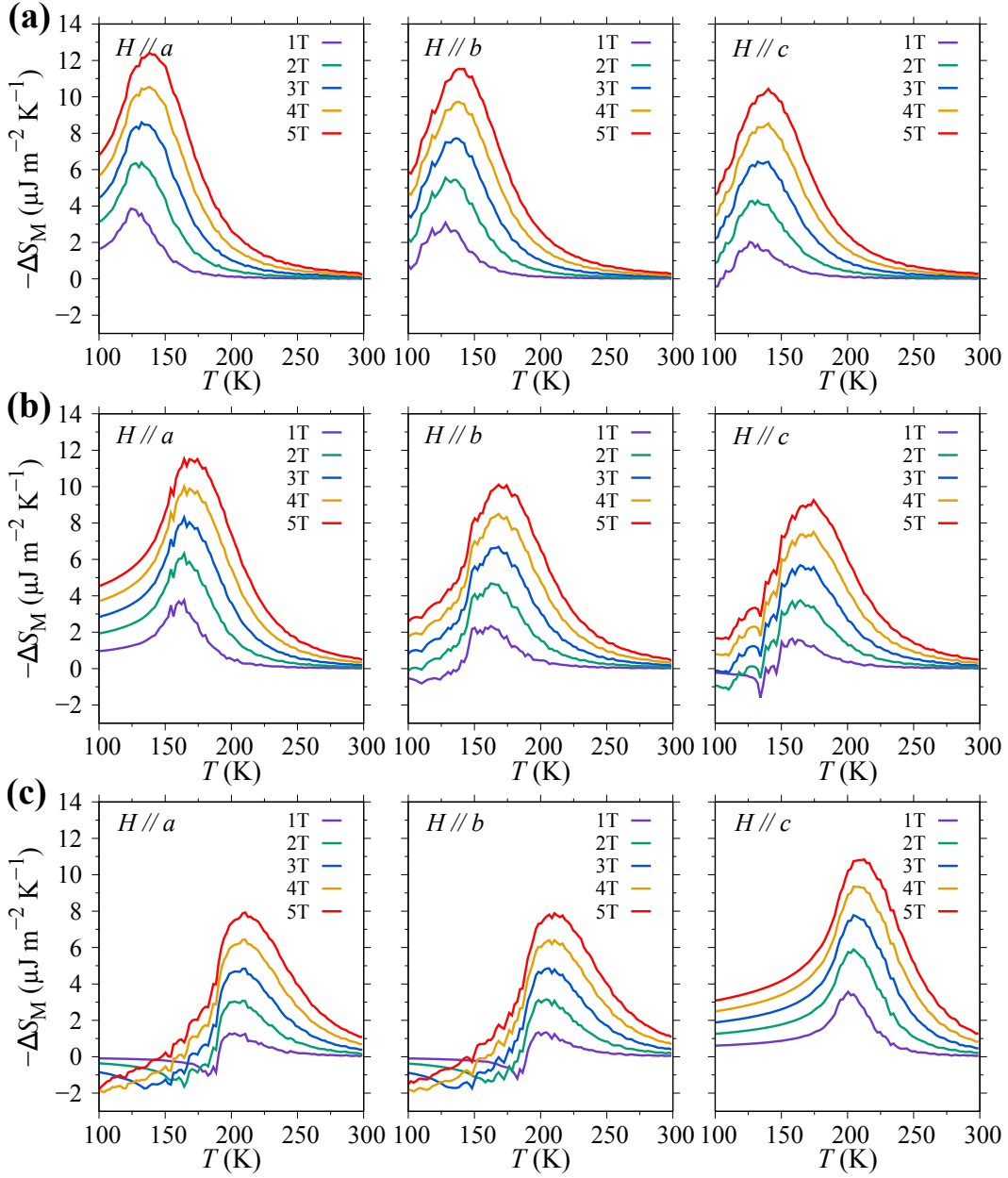

Fig. S16. Temperature dependences of $-\Delta S_\mathrm{M}$ of monolayer (a) CrSeCl, (b) CrSeBr and (c) CrSeI for $H \parallel a$, $H \parallel b$, and $H \parallel c$ in different magnetic fields, respectively.

17

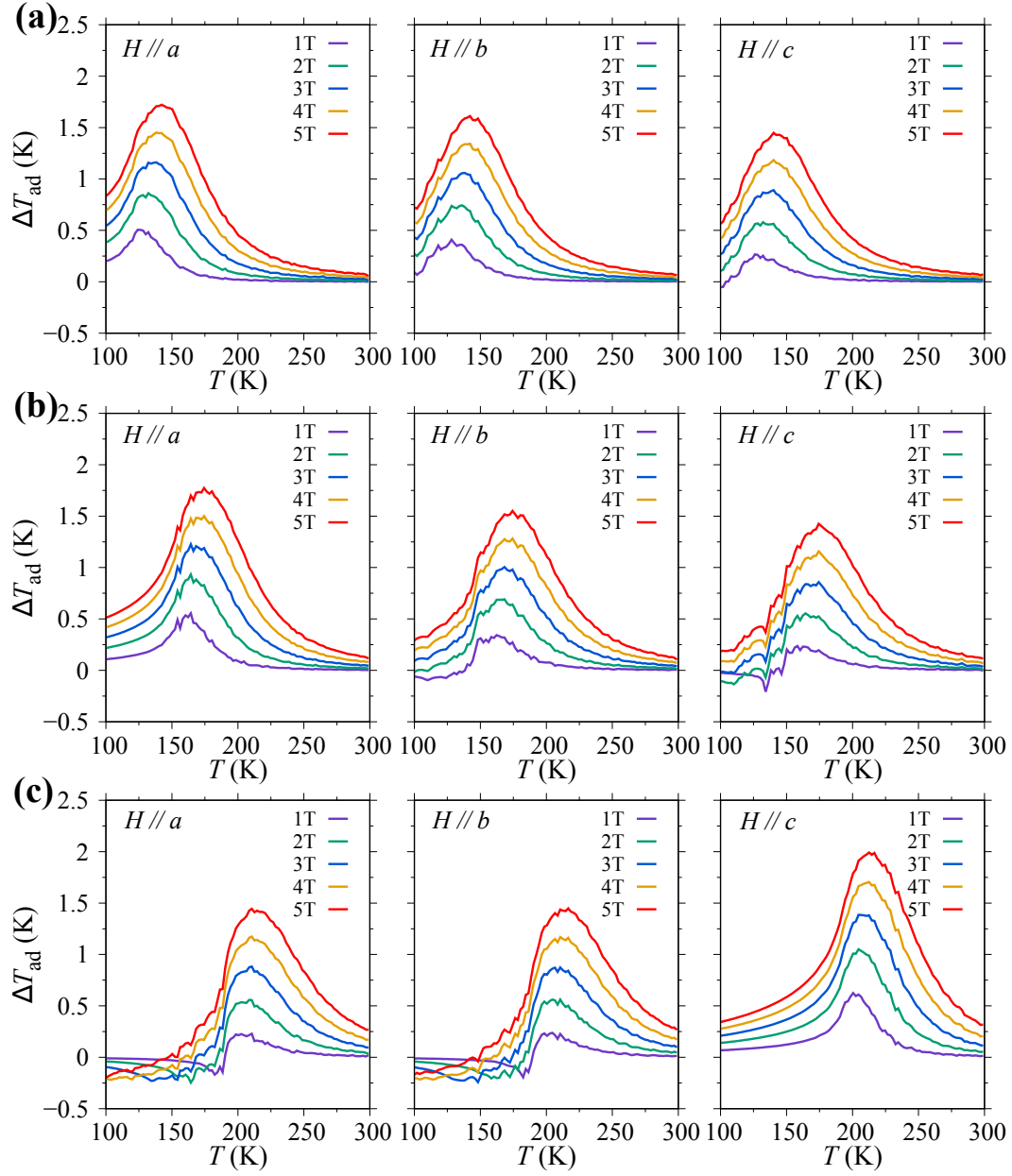

Fig. S17. Temperature dependences of $\Delta T_{\text{ad}}$ of monolayer (a) CrSeCl, (b) CrSeBr and (c) CrSeI for $H \parallel a$, $H \parallel b$, and $H \parallel c$ in different magnetic fields, respectively.

18

Fig. S18. Magnetic-field dependences of (a) $-\Delta S_\mathrm{M}^\mathrm{max}$ and (b) $\Delta T_\mathrm{ad}^\mathrm{max}$ of monolayers.

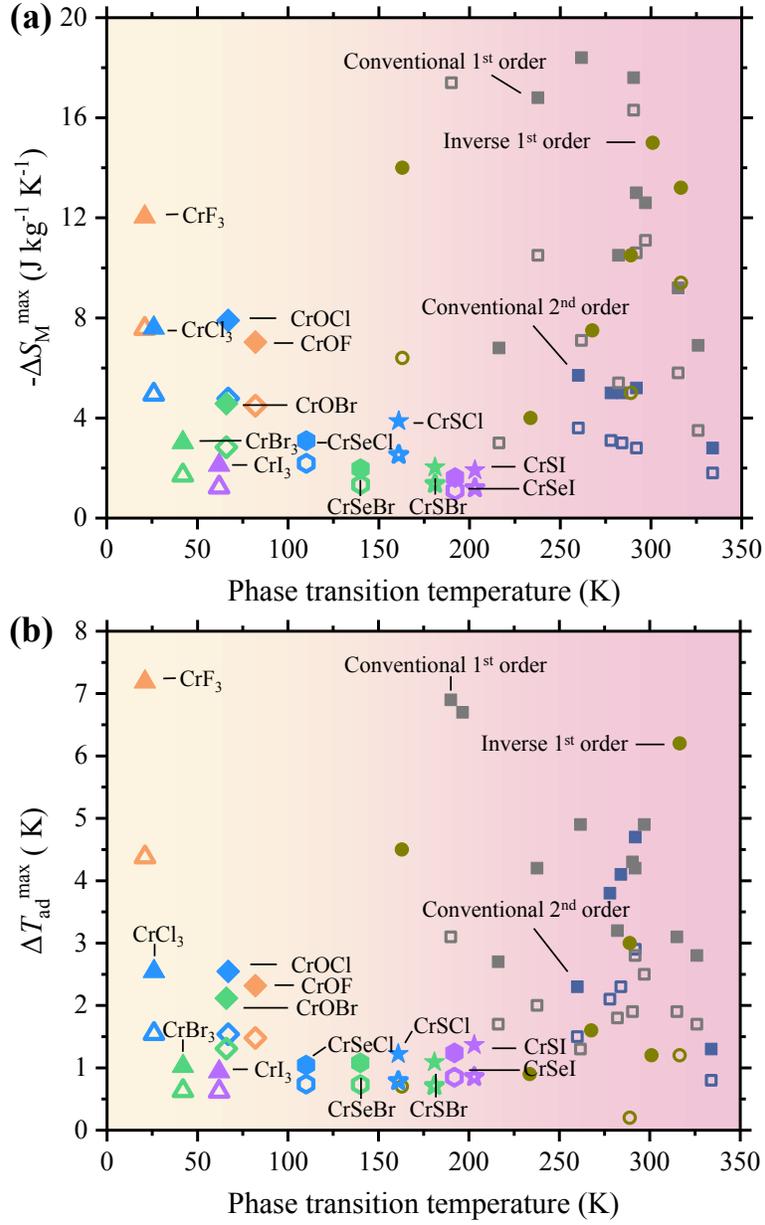

Fig. S19. The theoretical and ideal comparison of (a) $-\Delta S_\mathrm{M}^\mathrm{max}$ and (b) $\Delta T_\mathrm{ad}^\mathrm{max}$ between the monolayers and classical bulk magnetocaloric materials under low magnetic field (i.e., conventional 1st order materials, conventional 2nd order materials, inverse 1st order materials[5]). The hollow and solid points represent the data under 1 T and 2 T, respectively.

20

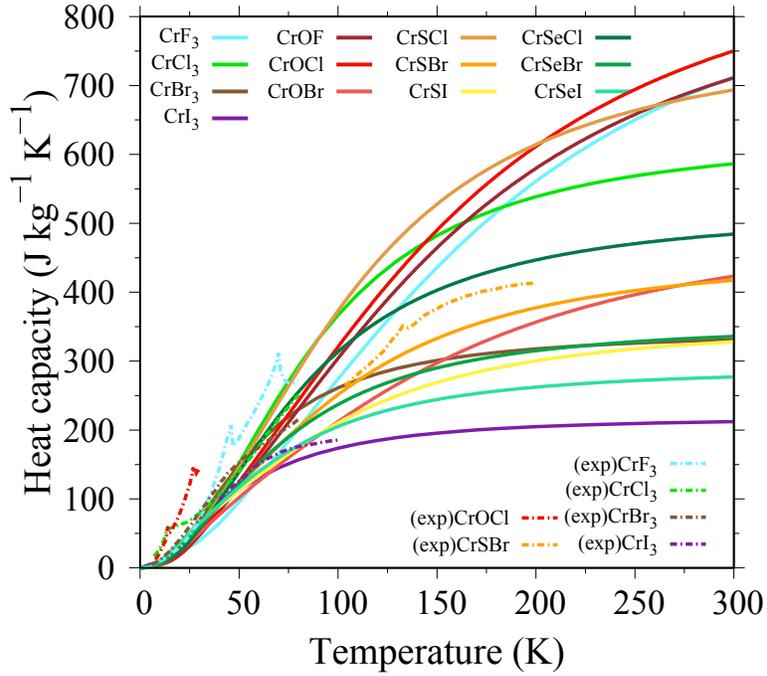

Fig. S20. Temperature dependences of heat capacity acquired by phonon calculations. The dashed lines are from previous bulk materials experimental measurements. The experimental data of single crystal of bulk $CrF_3$,[6] $CrCl_3$,[6] $CrBr_3$,[7] $CrI_3$,[3] $CrOCl$[8] and $CrOBr$[9] is drawn with dash lines.

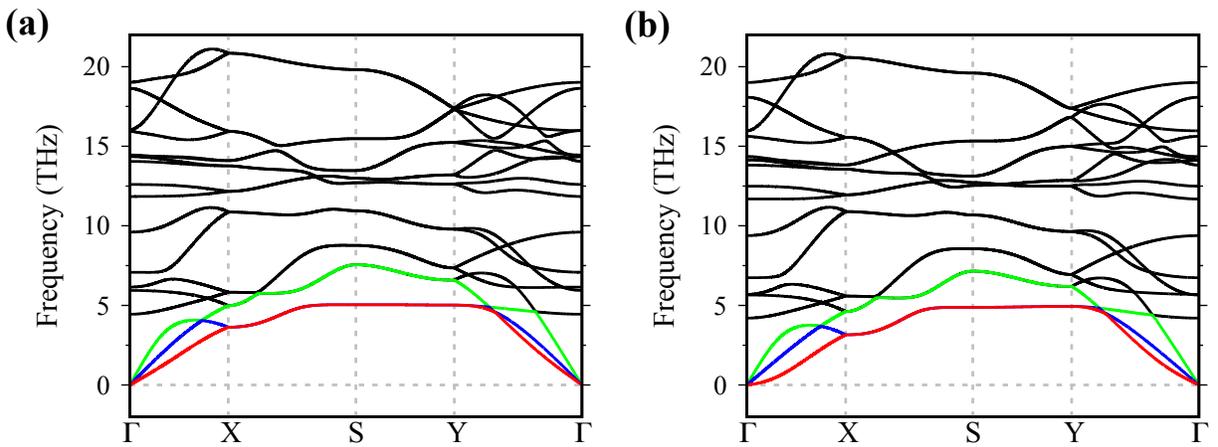

Fig. S21. Phonon dispersion spectra of monolayer CrOF under the 5% (a) $a$-axis and (b) $b$-axis compressive strain.

# References

(1) Sadhukhan, B.; Bergman, A.; Kvashnin, Y. O.; Hellsvik, J.; Delin, A. Spin-lattice couplings in two-dimensional CrI$_3$ from first-principles computations. *Physical Review B* **2022**, *105*, 104418.

(2) Staros, D.; Hu, G.; Tiihonen, J.; Nanguneri, R.; Krogel, J.; Bennett, M. C.; Heinonen, O.; Ganesh, P.; Rubenstein, B. A combined first principles study of the structural, magnetic, and phonon properties of monolayer CrI$_3$. *The Journal of Chemical Physics* **2022**, *156*, 014707.

(3) Liu, Y.; Petrovic, C. Anisotropic magnetocaloric effect in single crystals of CrI$_3$. *Physical Review B* **2018**, *97*, 174418.

(4) Deng, Y.; Yu, Y.; Song, Y.; Zhang, J.; Wang, N. Z.; Sun, Z.; Yi, Y.; Wu, Y. Z.; Wu, S.; Zhu, J.; Wang, J.; Chen, X. H.; Zhang, Y. Gate-tunable room-temperature ferromagnetism in two-dimensional Fe$_3$GeTe$_2$. *Nature* **2018**, *563*, 94–99.

(5) Gottschall, T.; Skokov, K. P.; Fries, M.; Taubel, A.; Radulov, I.; Scheibel, F.; Benke, D.; Riegg, S.; Gutfleisch, O. Making a cool choice: The materials library of magnetic refrigeration. *Advanced Energy Materials* **2019**, *9*, 1970130.

(6) Hansen, W. N.; Griffel, M. Heat capacities of CrF$_3$ and CrCl$_3$ from 15 to 300 K. *The Journal of Chemical Physics* **1958**, *28*, 902–907.

(7) Yu, X.; Zhang, X.; Shi, Q.; Tian, S.; Lei, H.; Xu, K.; Hosono, H. Large magnetocaloric effect in van der Waals crystal CrBr$_3$. *Frontiers of Physics* **2019**, *14*, 6–10.

(8) Zhang, T. et al. Magnetism and optical anisotropy in van der Waals antiferromagnetic insulator CrOCl. *ACS Nano* **2019**, *13*, 11353–11362.

(9) Lee, K.; Dismukes, A. H.; Telford, E. J.; Wiscons, R. A.; Wang, J.; Xu, X.; Nuckolls, C.;

Dean, C. R.; Roy, X.; Zhu, X. Magnetic order and symmetry in the 2D semiconductor CrSBr. *Nano Letters* **2021**, *21*, 3511–3517.


\end{document}